\documentclass[aps,prd,floatfix,amsmath,amssymb]{revtex4}
\usepackage{graphicx}
\usepackage[toc,page]{appendix}

\allowdisplaybreaks

\DeclareMathAlphabet{\mathpzc}{OT1}{pzc}{m}{it}
\makeatletter
\newbox\slashbox \setbox\slashbox=\hbox{$/$}
\newbox\Slashbox \setbox\Slashbox=\hbox{\large$/$}
\def\pFMslash#1{\setbox\@tempboxa=\hbox{$#1$}
  \@tempdima=0.5\wd\slashbox \advance\@tempdima 0.5\wd\@tempboxa
  \copy\slashbox \kern-\@tempdima \box\@tempboxa}
\def\pFMSlash#1{\setbox\@tempboxa=\hbox{$#1$}
  \@tempdima=0.5\wd\Slashbox \advance\@tempdima 0.5\wd\@tempboxa
  \copy\Slashbox \kern-\@tempdima \box\@tempboxa}

\def\miss#1{\ifmmode{/\mkern-11mu #1}\else{${/\mkern-11mu #1}$}\fi}
\makeatother

\begin{document}
\title{$WWV$ $(V=\gamma,Z)$ vertex in the Georgi-Machacek model}
\author{M. A. Arroyo-Ure\~na}
\author{G. Hern\'andez-Tom\'e}
\author{G. Tavares-Velasco}
 \affiliation{Facultad de Ciencias F\'{i}sico Matem\'aticas, Benem\'erita Universidad Au\-t\'o\-no\-ma de Puebla,
 Apartado postal 1152, 72001 Puebla, Pue., M\'exico.}

\begin{abstract}
The CP-even static form factors $\Delta\kappa'_V$ and $\Delta Q_V$ ($V=\gamma,\, Z$) associated with
the $WWV$ vertex are studied in the context of the Georgi-Machacek model (GMM), which predicts nine
new scalar bosons accommodated in a singlet, a triplet and a fiveplet. General expressions for the
one-loop contributions  to $\Delta\kappa'_V$ and
$\Delta Q_V$ arising from neutral, singly and doubly charged scalar bosons   are obtained in terms
of both parametric integrals and Passarino-Veltman scalar functions, which can be numerically
evaluated. It is found that the GMM yields 15 (28) distinct contributions to $\Delta\kappa'_\gamma$
and $\Delta Q_\gamma$ ($\Delta\kappa'_Z$ and $\Delta Q_Z$), though several of them are naturally
suppressed. A numerical analysis is done in the region of parameter space  still consistent with
current experimental data and  it is found that the largest contributions to $\Delta\kappa'_V$
arise from Feynman diagrams with two nondegenerate scalar bosons in the loop, with
values of the order of $a=g^2/(96\pi^2)$ reached when there is a large splitting between the masses
of these scalar bosons. As for $\Delta Q_V$, it reaches values as large as $10^{-2}a$ for the
lightest allowed scalar bosons, but it decreases rapidly as one of the masses of the scalar bosons
becomes large. Among the new contributions of the GMM to the $\Delta\kappa'_V$ and $\Delta Q_V$ form
factors are those induced by the $H_5^\pm W^\mp Z$  vertex, which arises at the tree-level  and
is a unique prediction of this model.
\end{abstract}

 \date{\today}
\preprint{}
\maketitle

\section{Introduction}
\label{Intro}
The observation of a 125 GeV Higgs-like particle by the CMS \citep{Chatrchyan:2012xdj} and ATLAS \citep{Aad:2012tfa} collaborations hints that  the Higgs mechanism, responsible for mass generation of elementary particles, is realized in nature. So far, the current measurements of this  particle's properties are consistent with the standard model (SM) Higgs boson. However, a more detailed and precise analysis is still necessary to confirm whether this particle is the SM Higgs boson or any other  remnant scalar boson arising in an extended scalar sector from a scenario beyond the SM. In fact, from a theoretical point of view, there is no  fundamental reason for a minimal Higgs sector, as occurs in the SM. It is therefore appropriate to consider additional scalar representations, which could have a role in the symmetry breaking mechanism and establish a relationship with a yet undiscovered sector.

Despite the great success of the SM, several extension models have been conjectured in order to solve the puzzle of some of the questions still unanswered by this theory.  In this context, models with  scalar triplet representations have attracted considerable attention due to their appealing  features, such as the possibility of implementing the seesaw mechanism to endow the neutrinos with naturally light Majorana masses (the so called type-II seesaw),  the appearance of the $H^{\pm}W^{\mp}Z$  coupling at the tree level, and the presence of doubly charged scalar particles. In this respect,  the Georgi-Machacek model (GMM)  \citep{Georgi:1985nv,Gunion:1989ci} is one of the most attractive Higgs triplet models  as it preserves the relationship  $\rho=1$  at the tree level via  an $SU(2)$ custodial symmetry. The GMM is based mainly on the SM but in the scalar sector introduces a complex scalar triplet  $\chi$, a real scalar triplet $\xi$, and the usual complex scalar doublet $\phi$ under the $SU(2)_L\times U(1)_Y$ gauge symmetry. After the spontaneous symmetry breaking, the physical scalar spectrum of the GMM is given by the SM-like Higgs boson $h$ and one extra CP-even singlet $H$, one scalar triplet $H_3$ ($H_3^0$, $H_3^\pm$), and one scalar fiveplet $H_5$ ($H_5^0$, $H_5^{\pm\pm}$, $H_5^{\pm}$). All of these multiplets are mass degenerate as a result of the custodial symmetry. The phenomenology of the GMM has been broadly studied over the recent years \citep{Delgado:2016arn, Godunov:2015lea, Chiang:2015kka, Hartling:2014xma, Godunov:2014waa, Chiang:2014hia, Englert:2013zpa, Godfrey:2010qb, Logan:2015xpa, Hartling:2014zca, Chiang:2012cn, Degrande:2015xnm, Chiang:2015amq, Chiang:2015rva}.  For instance, a study of the search and production of the GMM Higgs bosons at the LHC has been analyzed in \citep{Chiang:2015amq,Degrande:2015xnm}, and its phenomenology at a future electron-positron collider has been reported in \citep{Chiang:2015rva}.

Even if there is not enough energy available to produce the new scalar particles predicted by the GMM, one can search for  their virtual effects through some observables.
Particular interest has been put on the radiative corrections to the $WWV$ $(V=\gamma, Z)$ vertex, which represents a very sensitive scenario to search for any NP effects and test the gauge sector of the SM. In fact, the  one-loop corrections to the on-shell $WW\gamma$ vertex,  which define the static electromagnetic properties of the $W$ gauge boson,  was one of the first ever one-loop calculations within the SM \citep{Bardeen:1972vi}, followed by a plethora of calculations of the respective contributions of several SM  extensions, such as the
two-Higgs doublet model (THDM) \citep{Couture:1987eu}, the minimal supersymmetric standard model (MSSM) \citep{Lahanas:1994dv}, left-right symmetric theories \citep{Larios:1995rz}, extra dimensions \citep{FloresTlalpa:2010rm}, the littlest Higgs model \citep{Moyotl:2010ss}, 331 models \citep{Montano:2005gs,GarciaLuna:2003tj}, effective theories \cite{HernandezSanchez:2006sw,TavaresVelasco:2004gx,TavaresVelasco:2003ka}, etc. In contrast with the on-shell $WW\gamma$ vertex, additional difficulties in the calculation of the on-shell  $WWZ$ vertex arise due to the nonzero mass of the  $Z$ gauge boson. In this respect, the study of radiative corrections to the $WWZ$ vertex has been the  focus of attention when the $Z$ boson is off-shell as can be found in Refs. \citep{Argyres:1995ib, RamirezZavaleta:2007hm, Montano:2005gs, FloresTlalpa:2010rm}.  This type of calculations are in general gauge dependent and require special techniques, such as the pinch technique, to extract the  relevant  physical information.

The on-shell $WWV$ vertex can be written in terms of four form factors that define the CP-even and CP-odd static properties of the $W$ boson. The two CP-odd form factors $\Delta\widetilde\kappa '_{V}$ and $\Delta\widetilde Q_{V}$ are absent up to the one-loop level in the SM and are thus expected to be negligibly small.  As far as the CP-even form factors $\Delta\kappa '_{V}$ and $\Delta Q_{V}$ are concerned, they arise at the one-loop level in the SM and any other renormalizable theory, thereby being highly sensitive to NP effects.

The most general dimension-4 $CP$-conserving $WWV$ ($V=\gamma, Z$)  vertex is given by \cite{Hagiwara:1986vm}

\begin{equation}
\mathcal{L}=-ig_{V}\left\{ g_{1}^{V}V_{\mu}\left(W^{-\mu\nu}W_{\nu}^{+}-W^{+\mu\nu}W_{\nu}^{-}\right)+\kappa_{V}V_{\mu\nu}W^{+\mu}W^{-\nu}+\frac{\lambda_{V}}{M_{W}^{2}}V^{\mu\nu}W_{\nu}^{+\alpha}W_{\alpha\mu}^{-}\right\},\label{WWV-lagrangian}
\end{equation}
where $g_V$ stand for the $WWV$  tree-level coupling constant (in the SM $g_\gamma=gs_W$ and $g_Z=gc_W$). Here $g_{1}^{V}$,  $\kappa_{V}$  and $\lambda_{V}$  represent form factors that can receive radiative corrections. In the SM,  $SU(2)_{L}\times U(1)$  gauge symmetry implies $g_{1}^{V}=\kappa_{V}=1$  and $\lambda_{V}=0$ at the tree level.

\begin{figure}[htb!]
\begin{center}
\includegraphics[width=7.5cm]{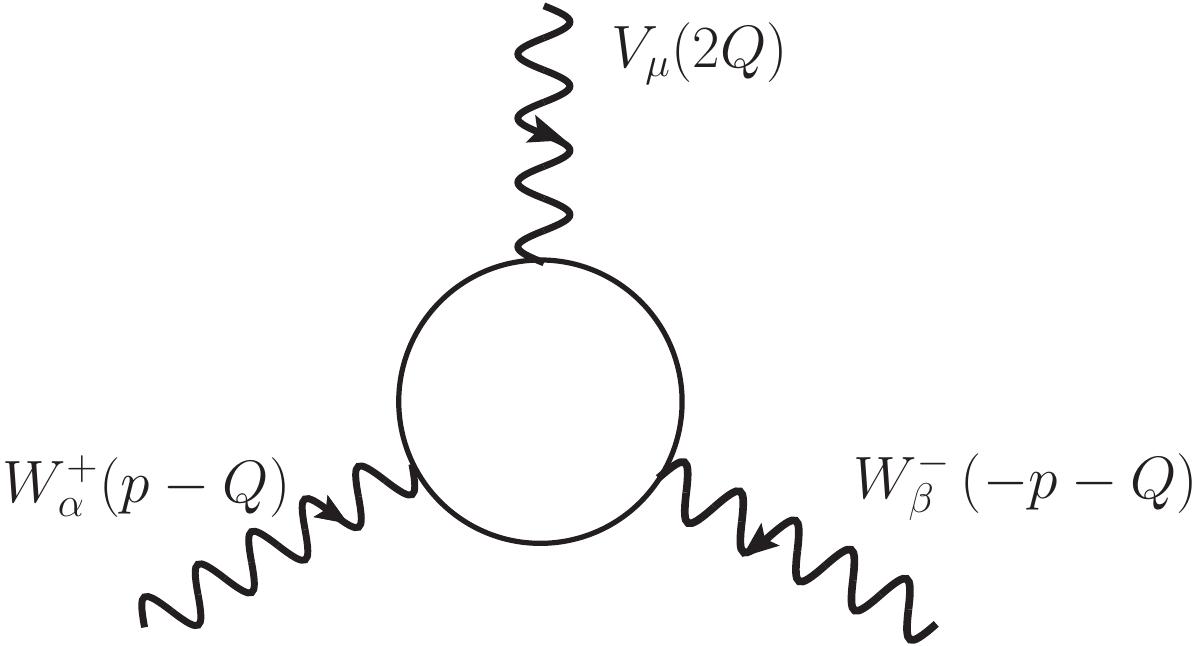}
\caption{Nomenclature for the $WWV$ vertex function. The circle denotes   radiative contributions.} \label{WWV-vertex}
\par\end{center}
\end{figure}

The vertex function that determines the $WWV$  coupling can be written as

\begin{eqnarray}
\Gamma^{\mu\alpha\beta}_{V}&=&ig_V\Bigg\{{A\left[2p^{\mu}g^{\alpha\beta}+4\left(Q^{\beta}g^{\mu\alpha}-Q^{\alpha}g^{\mu\beta}\right)\right]}
\nonumber\\&&+2\Delta \kappa'_V\left(Q^{\beta}g^{\mu\alpha}-Q^{\alpha}g^{\mu\beta}\right)+\frac{4\Delta Q_V}{m_W^2}\left(p^{\mu}Q^{\alpha}Q^{\beta}-\frac{1}{2}m_V^2 p^{\mu}g^{\alpha\beta} \right)\Bigg\}, \label{vertex-function}
\end{eqnarray}
where we have used the convention employed in \cite{Bardeen:1972vi} for the external momenta, as shown in Fig. \ref{WWV-vertex}.
The form factors defined in Eq. (\ref{vertex-function}) are related to those appearing in Eq. (\ref{WWV-lagrangian}) according to

\begin{eqnarray}
\Delta\kappa'_V&\equiv& \kappa_V-1+\lambda_V,\label{Deltakp}\\ \qquad \Delta Q_V&\equiv& -2\lambda_V\label{DeltaQ}.
\end{eqnarray}
It is worth mentioning that the  definition $\Delta\kappa_V= \kappa_V-1$  is customarily  used in experimental works, where the constraints are given traditionally as bounds on $\Delta\kappa_V$ and $\lambda_V$, whereas in theoretical works it has been usual to present the analytical results  in terms of $\Delta\kappa'_V$ and $\Delta Q_V$.


For the photon, $\kappa_\gamma$ and $\lambda_\gamma$ are related to the  magnetic dipole moment $\mu_W$ and the
electric quadrupole moment $Q_W$ of the $W$ gauge boson as follows

\begin{equation}
\mu_W=\frac{e}{2\,m_W}\,\left(1+\kappa_\gamma+\lambda_\gamma\right),
\end{equation}

\begin{equation}
Q_W=-\frac{e}{m_W^2}\,\left(\kappa_\gamma-\lambda_\gamma\right).
\end{equation}

In this work, we will calculate the contributions of the complete scalar sector of the GMM to  the $\Delta\kappa '_{V}$ and $\Delta Q_{V}$ form factors, which could be at the reach of the  future linear collider  experiments \citep{Moortgat-Picka:2015yla, Bian:2015zha}. The structure of our work is organized as follows.  An overview of the GMM is presented in Section \ref{Model}. In Sec. \ref{FormFactors} we present the analytical expressions for the $\Delta\kappa '_V$ and $\Delta Q$ form factors, whereas the numerical results are analyzed in Sec. \ref{NumericalDiscussion} and the conclusions and outlook are presented in Sec. \ref{Conclusions}.

\section{The Georgi-Machacek Model}
\label{Model}

The scalar sector of the GMM is composed by an isospin complex triplet $\chi$ with  hypercharge  $Y = 2$, a real triplet $\xi$ with $Y = 0$, and the usual SM isospin doublet $\phi$ with $Y = 1$. The global $SU(2)_{L}\times SU(2)_R$ custodial symmetry is manifest  by writing the fields as

\begin{equation}
\Phi=\left(\begin{array}{cc}
\phi^{0*} & \phi^{+}\\
-\phi^{+*} & \phi^{0}
\end{array}\right), \qquad
X=\left(\begin{array}{ccc}
\chi^{0*} & \xi^{+} & \chi^{++}\\
-\chi^{+*} & \xi^{0} & \chi^{+}\\
\chi^{++*} & -\xi^{+*} & \chi^{0}
\end{array}\right),\label{fields}
\end{equation}
where $\Phi$ and $X$ transform under the custodial symmetry as $\Phi\to U_L\Phi U_R^{\dagger}$ and  $X \to U_L X U_R^{\dagger}$ with $U_{L,R}=e^{(i\theta^a_{L,R}T^a)}$. Here $T^a=t^a$ stands for the $SU(2)$ generators in the triplet representation

\begin{equation}
t^{1}=\frac{1}{\sqrt{2}}\left(\begin{array}{ccc}
0 & 1 & 0\\
1 & 0 & 1\\
0 & 1 & 0
\end{array}\right),\qquad t^{2}=\frac{1}{\sqrt{2}}\left(\begin{array}{ccc}
0 & -i & 0\\
i & 0 & -i\\
0 & i & 0
\end{array}\right),\qquad t^{3}=\frac{1}{\sqrt{2}}\left(\begin{array}{ccc}
1 & 0 & 0\\
0 & 0 & 0\\
0 & 0 & -1
\end{array}\right),
\end{equation}
whereas for the doublet representation $T^a=\sigma^a/2$, with $\sigma^a$ the Pauli matrices.

The neutral members of the fields in Eq. (\ref{fields}) develop  a nonzero vacuum expectation value (VEV) defined by $\langle\Phi\rangle=\frac{v_\phi}{\sqrt{2}}I_{2\times2} $ and $\langle X \rangle=v_{\chi} I_{3\times3}$, with $I_{n\times n}$ the $n\times n$ identity matrix. The masses of the $W$ and $Z$ gauge bosons constrain the VEV values as follow

\begin{equation}
v_{\phi}^2+8v_{\chi}^2\equiv v^2 =\frac{1}{\sqrt{2}G_F}\approx(246 \textrm{ GeV})^2.
\end{equation}
The kinetic  Lagrangian of the scalar sector, out of which  the gauge boson masses arise, takes the form

\begin{equation}
\mathcal{L}=\frac{1}{2}{\rm Tr}\left[\left(D_{\mu}\Phi\right)^{\dagger}\left(D^{\mu}\Phi\right)\right]+\frac{1}{2}{\rm Tr}\left[\left(D_{\mu}X\right)^{\dagger}\left(D^{\mu}X\right)\right],
\label{Kinetic_Lagrangian}
\end{equation}
with the covariant derivative given by

\begin{eqnarray}
D_{\mu}\Phi&=&\partial_{\mu}\Phi+i\frac{g}{2}\tau^a W^a_\mu \Phi-i\frac{g'}{2}\tau_{3}B_\mu,
\end{eqnarray}
and a similar expression for $D_{\mu}X$. As for the most general scalar potential  that obeys the custodial symmetry, it can be written as

\begin{eqnarray}
V\left(\Phi,X\right) & = & \frac{\mu_{2}^{2}}{2}\textrm{Tr}\left(\Phi^{\dagger}\Phi\right)+\frac{\mu_{3}^{2}}{2}\textrm{Tr}\left(X^{\dagger}X\right)+\lambda_{1}\left[\textrm{Tr}\left(\Phi^{\dagger}\Phi\right)\right]^{2}+\lambda_{2}\textrm{Tr}\left(\Phi^{\dagger}\Phi\right)\textrm{Tr}\left(X^{\dagger}X\right)\\
 &  & +\lambda_{3}\textrm{Tr}\left(X^{\dagger}XX^{\dagger}X\right)+\lambda_{4}\left[\textrm{Tr}\left(X^{\dagger}X\right)\right]^{2}-\lambda_{5}\textrm{Tr}\left(\Phi^{\dagger}\tau^{a}\Phi\tau^{b}\right)\textrm{Tr}\left(X^{\dagger}t^{a}Xt^{b}\right)\\
 &  & -M_{1}\textrm{Tr}\left(\Phi^{\dagger}\tau^{a}\Phi\tau^{b}\right)\left(UXU^{\dagger}\right)_{ab}-M_{2}\textrm{Tr}\left(X^{\dagger}t^{a}Xt^{b}\right)\left(UXU^{\dagger}\right)_{ab},\label{potencial}
\end{eqnarray}
where the matrix $U$, which rotates $X$ into the Cartesian basis, is given by

\begin{equation}
U=\left(\begin{array}{ccc}
-\frac{1}{\sqrt{2}} & 0 & \frac{1}{\sqrt{2}}\\
-\frac{i}{\sqrt{2}} & 0 & -\frac{i}{\sqrt{2}}\\
0 & 1 & 0
\end{array}\right).
\end{equation}

In order to obtain the physical scalar spectrum after the spontaneous symmetry breaking, it is appropriate to decompose the neutral fields into the real and imaginary parts in the following way

\begin{equation}
\phi^0\to \frac{v_{\phi}}{\sqrt{2}}+\frac{\phi^{0,r}+i\phi^{0,i}}{\sqrt{2}}, \quad \chi^0\to v_{\chi}+\frac{\chi^{0,r}+i\chi^{0,i}}{\sqrt{2}}, \quad \xi^0\to v_{\chi}+\xi^0.
\end{equation}

The physical fields are organized by their transformation properties under the $SU(2)$ custodial symmetry into a fiveplet, a triplet, and two singlets. The fiveplet and triplet states are given by

\begin{equation}
H_{5}^{++}=\chi^{++},\qquad H_{5}^{+}=\frac{1}{\sqrt{2}}\left(\chi^{+}-\xi^{+}\right),\qquad H_{5}^{0}=-\sqrt{\frac{2}{3}}\xi^{0}+\sqrt{\frac{1}{3}}\chi^{0,r},
\end{equation}
\begin{equation}
H_{3}^{+}=-s_{H}\phi^{+}+\frac{c_{H}}{\sqrt{2}}\left(\chi^{+}+\xi^{+}\right),\qquad H_{3}^{0}=-s_{H}\phi^{0,i}+c_{H}\chi^{0,i},
\end{equation}
where the mix between $v_{\phi}$ and $v_{\chi}$ is parametrized in terms of a mixing angle $\theta_H$ according to

\begin{equation}
c_H\equiv \cos \theta_H=\frac{v_{\phi}}{v}, \qquad s_H\equiv \sin \theta_H=\frac{2\sqrt{2}v_{\chi}}{v}.
\end{equation}

The two singlet mass eigenstates are given by

\begin{equation}
h=\cos\alpha\phi^{0,r}-\sin\alpha H_{1}^{0'},\qquad H=\sin\alpha\phi^{0,r}+\cos\alpha H_{1}^{0'},
\end{equation}
where $H_{1}^{0'}=\sqrt{\frac{1}{3}}\xi^{0}+\sqrt{\frac{2}{3}}\chi^{0,r}$, whereas $h$ is associated with the  SM Higgs boson. The mixing angle $\alpha$ is given  by

\begin{equation}
\sin2\alpha=\frac{2\mathcal{M}_{12}^{2}}{m_{H}^{2}-m_{h}^{2}},
\end{equation}
with

\begin{equation}
\mathcal{M}_{12}^{2}=\frac{\sqrt{3}}{2}v_{\phi}\left[-M_{1}+4\left(2\lambda_{2}-\lambda_{5}\right)v_{\chi}\right].
\end{equation}

A peculiarity  of this model is that the $H_{5}$ states are fermiophobic, which stems from the fact that there is no doublet field in the custodial fiveplet. As far as the masses for the fiveplet and triplet are concerned, they are degenerate at the tree
level and are expressed in terms of the respective VEVs and the parameters involved in the scalar potential as follows

\begin{equation}
m_{5}^{2}=\frac{M_{1}}{4v_{\chi}}v_{\phi}^{2}+12M_{2}v_{\chi}+\frac{3}{2}\lambda_{5}v_{\phi}^{2}+8\lambda_{3}v_{\chi}^{2},
\end{equation}

\begin{equation}
m_{3}^{2}=\frac{M_{1}}{4v_{\chi}}\left(v_{\phi}^{2}+8v_{\chi}^{2}\right)+\frac{\lambda_{5}}{2}\left(v_{\phi}^{2}+8v_{\chi}^{2}\right)=\left(\frac{M_{1}}{4v_{\chi}}+\frac{\lambda_{5}}{2}\right)v^{2}.
\end{equation}
On the other hand, the  singlet masses are given by
\begin{equation}
m_{h,H}^{2}=\frac{1}{2}\left[\mathcal{M}_{11}^{2}+\mathcal{M}_{22}^{2}\mp\sqrt{\left(\mathcal{M}_{11}^{2}-\mathcal{M}_{22}^{2}\right)^{2}+4\left(\mathcal{M}_{12}^{2}\right)^{2}}\right],
\end{equation}
with

\begin{equation}
\mathcal{M}_{11}^{2}=8\lambda_{1}v_{\phi}^{2},
\end{equation}
and
\begin{equation}
\mathcal{M}_{22}^{2}=\frac{M_{1}v_{\phi}^{2}}{4v_{\chi}}-6M_{2}v_{\chi}+8\left(\lambda_{3}+3\lambda_{4}\right)v_{\chi}^{2}.
\end{equation}

From the kinetic Lagrangian (\ref{Kinetic_Lagrangian}) one can also obtain the interactions between the SM gauge bosons and all the new scalar bosons predicted by the GMM. The full set of Feynman rules  can be found in Refs. \citep{Gunion:1989ci, Chiang:2012cn}.  As far as our calculation is concerned, apart from the usual SM vertex of the type $W^-W^+ V$ ($V=\gamma,\,Z$), in the GMM the following new type of vertices can arise $\phi^\mp_A\phi^\pm_AV$,  $\phi^{\mp\mp}_A\phi^{\pm\pm}_AV$, $\phi^\mp_A\phi^{\pm\pm}_B W^\mp$, $\phi^\mp_A\phi^0_B W^\pm$, $\phi^\mp_A\tilde\phi^0_B W^\pm$, $W^\mp W^\mp\phi^{\pm\pm}_A$, where  $\phi^0_I=h,\,H,\, H_5^0$,  $\tilde \phi^0_I=H^0_3$, $\phi^\mp_I= H^\mp_3,\, H^\mp_5$,  and  $\phi^{\mp\mp}_I= H^{\mp\mp}_5$  ($I=A,\, B$). In addition, the $Z$ gauge boson has extra couplings of the form $\phi^\mp_A W^\pm Z$, $\phi_A^\mp \phi_B^\pm Z$, $\phi^0_A ZZ$, and  $\phi^0_A \tilde \phi_B^0 Z$. It turns out that all these vertices are just of three distinct types, namely, $X_A X_A V$ (three gauge bosons), $\phi_A \phi_B X_C$ (two scalar bosons and one gauge boson), and $\phi_A X_B X_C$ (one scalar boson and two gauge bosons), where $\phi_I$ ($I=A,\,B$) stands for a neutral,  singly charged or  doubly charged scalar boson, whereas $X_J$ ($J=A,\, B,\, C$) stands for a neutral or charged  gauge boson. Evidently, the allowed vertices  are dictated by electric charge conservation, Bose symmetry,  $CP$ invariance (as long as it is assumed to be conserved), etc.
However, the Lorentz structure is similar for each type of vertex  and so are the respective Feynman rules,  which  arise from the following Lagrangians:

\begin{equation}
\mathcal{L}_{X_A X_A V}=ig_{X_A X_A V}\left({X_A}_{\mu\nu}^{\dagger}V^{\nu}-{X_A}^{\mu\nu}{X_A}_{\mu}^{\dagger}V_{\nu}+V^{\mu\nu}{X_A}_{\mu}^{\dagger}{X_A}_{\nu}\right),
\label{Effective-Lagrangian1}
\end{equation}
\begin{equation}
\mathcal{L}_{\phi_A \phi_B X_C}=ig_{\phi_A\phi_BV}{X_C}^{\mu}\phi_{A}^{\dagger}\overleftrightarrow{\partial}\phi_B
,\label{Effective-Lagrangian2}
\end{equation}
and
\begin{equation}
\mathcal{L}_{X_A X_B \phi_C}= g_{X_AX_B\phi_C}X_A^{\mu}X_{B\mu}\phi_C,\label{Effective-Lagrangian3}
\end{equation}
with ${X_A}^{\mu\nu}=\partial_{\mu}{X_A}^{\nu}-\partial_{\nu}{X_A}^{\mu}$. We have assumed that $CP$ is conserved.

\begin{figure}[!htb]
\begin{center}
\includegraphics[width=15cm]{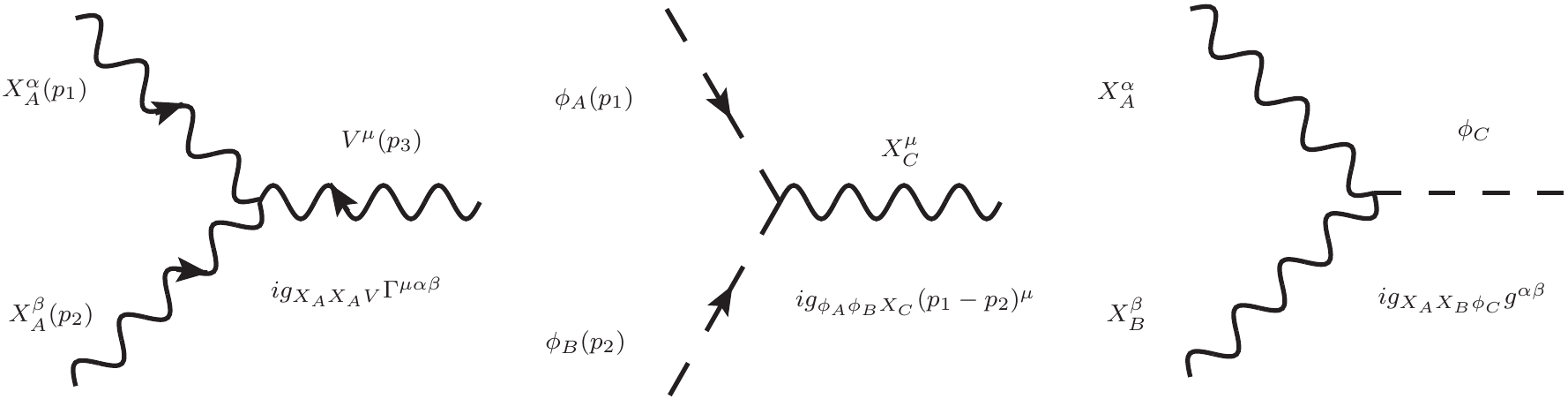}
\caption{Generic Feynman rules for the relevant vertices involved in our calculation. The arrows stand for the direction of the 4-momenta and $\Gamma_{\mu\alpha\beta}=g_{\mu\alpha}(p_1-p_3)_{\beta}+g_{\alpha\beta}(p_2-p_1)_{\mu}+g_{\beta\mu}(p_3-p_2)_{\alpha}$.
$V=\gamma, Z$, $\phi_{I}$ ($I=A,\,B,\,C$) denote a neutral singly, or doubly charged scalar boson, and $X_J$ ($J=A,\,B$) stands for a neutral or charged gauge boson. The electric charge and $CP$ properties of  the particles attached to each vertex are dictated by electric charge conservation,  Bose symmetry, $CP$ invariance, etc.
} \label{Feynman-Rules}
\par\end{center}
\end{figure}

For the photon, the only allowed vertices are $WW\gamma$, $\phi^\mp\phi^\pm \gamma$, and  $\phi^{\mp\mp}\phi^{\pm\pm} \gamma$, whereas the $Z$ gauge boson can also have nondiagonal couplings  to both charged and neutral scalar bosons.
From Eqs. (\ref{Effective-Lagrangian1}) - (\ref{Effective-Lagrangian3}),  generic  Feynman rules follow straightforwardly and are  shown in Fig. \ref{Feynman-Rules}. Therefore, we can perform a model-independent calculation and express our results in terms of   the coupling constants and the masses of the virtual particles.   In  particular, the coupling constants for the vertices allowed in the GMM  are presented in Appendix \ref{Coupling_Constants}.

\section{$\Delta\kappa'_V$ and $\Delta Q_V$ form factors in the GMM}
\label{FormFactors}

\begin{figure}[htb!]
\begin{center}
\includegraphics[width=15cm]{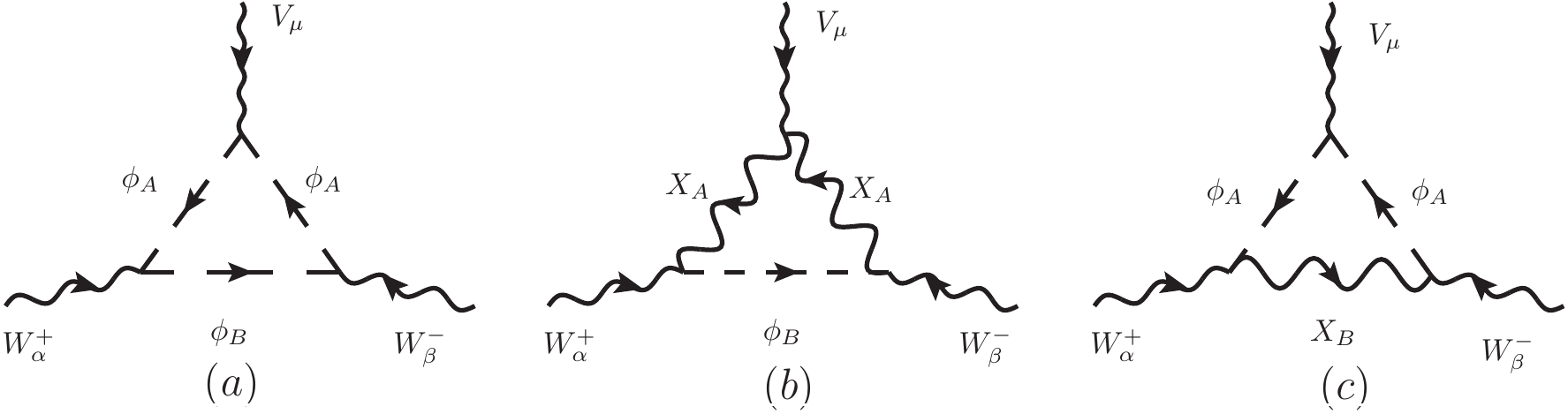}
\caption{ Generic Feynman diagrams for the new scalars contributions to both the $WW\gamma$ and  $WWZ$ vertices  involving only two distinct virtual particles. The arrows stand for the directions of the 4-momenta. The possible combinations of internal particles are given by the vertices allowed in each particular model. For instance, when $V=\gamma$,  the following electric charges of the internal particles are possible in the GMM,  in units of the positron charge: if $Q_A=-1$ then $Q_B=0$, if  $Q_A=1$ then $Q_B=2$, if $Q_A=-2$ then $Q_B=-1$.} \label{Diagrams-abc}
\par\end{center}
\end{figure}

We now turn to present the contributions of the scalar sector of the GMM to the $\Delta\kappa'_V$ and $\Delta Q_V$ form factors at the one-loop level.
In  this model, the new one-loop contributions   arise from  generic triangle diagrams (the bubble diagrams do not contribute) that can be classified according to the number of  distinct particles circulating into the loop.  In Fig. \ref{Diagrams-abc} we show a set of Feynman diagrams that  contribute to both the $WW\gamma$ and $WWZ$ vertices.  These  diagrams include just two distinct particles circulating inside the loop as they involve diagonal couplings of the form $\phi_A\phi_A V$ and $X_AX_A V$.

\begin{figure}[htb!]
\begin{center}
\includegraphics[width=8.5cm]{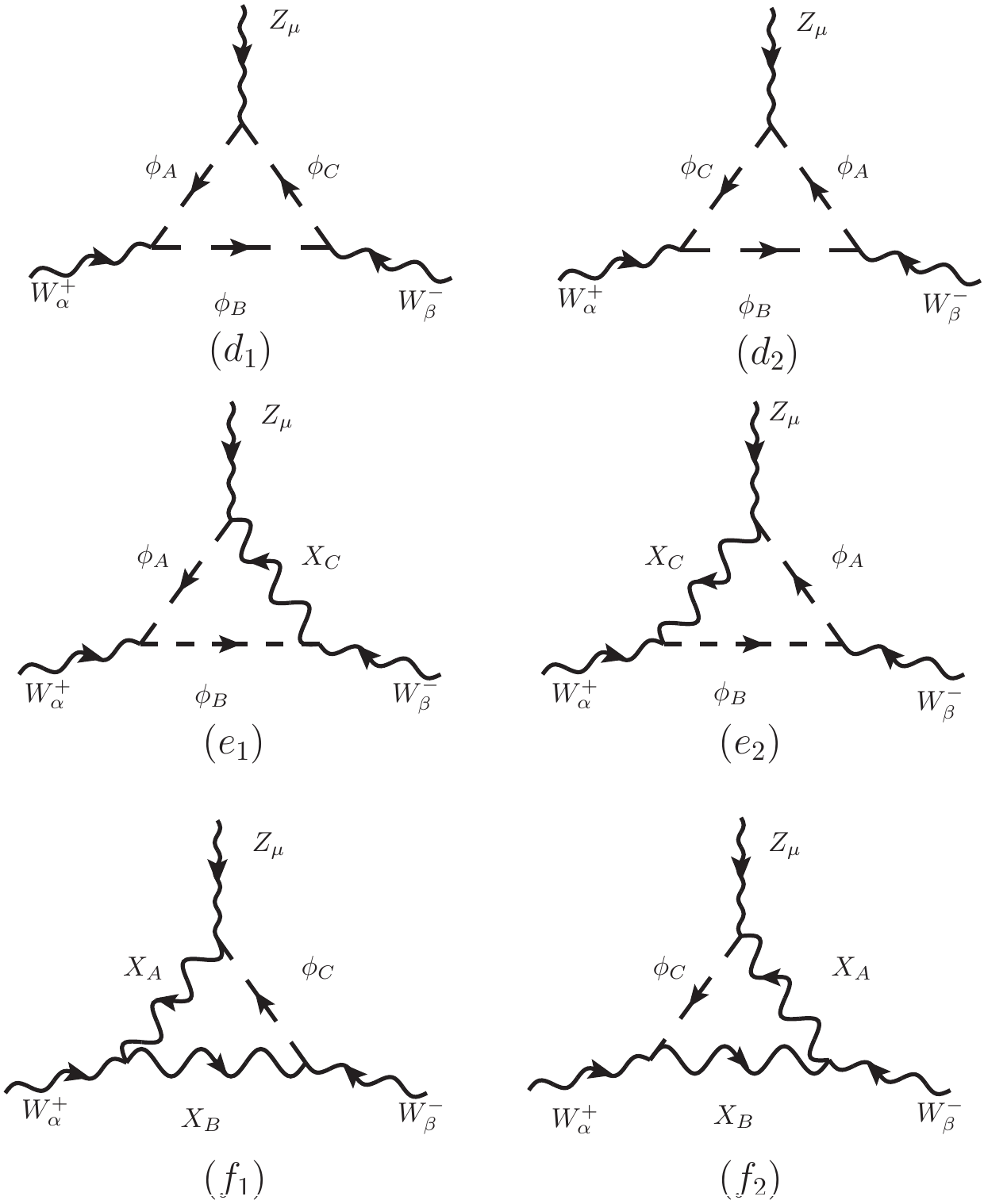}
\caption{Extra contributions to the $\Delta\kappa'_Z$ and $\Delta Q_Z$ form factors from nondiagonal couplings. As explained in  the text, the possible sets of internal particles  are determined by the vertices allowed in a particular model.} \label{Diagrams-def}
\end{center}
\end{figure}

Contrary to the couplings  of the photon to a pair of charged scalar bosons, which can  only be of diagonal type due to electromagnetic gauge invariance, the $Z$ gauge boson can have nondiagonal couplings to a pair of neutral or charged scalar bosons. Therefore, in addition  to the diagrams of Fig. \ref{Diagrams-abc}, the $\Delta\kappa'_Z$ and $\Delta Q_Z$ form factors can receive extra contributions from the Feynman diagrams  shown in Fig. \ref{Diagrams-def}, which  have three distinct particles circulating into the loop. Below we will present the contributions to  $\Delta\kappa'_V$ and $\Delta Q_V$   for all these types of diagrams.

Before presenting our results, some remarks about our calculation are in order:
\begin{itemize}

\item   The Feynman diagrams were evaluated via the unitary gauge. In order to make a cross check of our results we used both, the Feynman parametrization technique and the Passarino-Veltman method  to solve the loop integrals.

\item We verified that all  the   contributions of bubble diagrams to the $\Delta\kappa'_V$ and $\Delta Q_V$ form factors   involving  quartic vertices with two scalar bosons and two gauge bosons vanish, and thus the only contributions arise from  triangle diagrams.

  \item The mass shell and transversality conditions for the gauge bosons  enabled us to make the following replacements

\begin{equation}
 Q^2=\frac{m_V^2}{4},\qquad p\cdot Q=0,\qquad   p^2=m_W^2-\frac{m_V^2}{4}, \label{conditions1}
\end{equation}
and
\begin{equation}
p_{\alpha}\to Q_{\alpha}, \quad p_{\beta}\to-Q_{\beta}, \quad p_\mu\to 0, \label{conditions2}
\end{equation}
which results in a considerable simplification of the calculation.

\item  Instead of dealing with the calculation of the $WW\gamma$ and $WWZ$ vertices separately, we  performed instead the calculation of   the  general $WWV$ vertex, with $V$ a massive neutral gauge boson. We have exploited the fact that there are only  three generic trilinear vertices involved in the one-loop contributions to the $WWV$ vertex  and thus a   model independent calculation was done using the generic Feynman rules of Fig. \ref{Feynman-Rules}. The result for  the contribution of each type of  Feynman diagram will be presented in terms of loop functions, given as parametric integrals and also in terms of Passarino-Veltman scalar integrals, times a factor involving all the generic coupling constants associated with each vertex participating in the particular diagram. The contribution to the form factors of the  $WW\gamma$ and $WWZ$ vertices  follow easily from our general expressions after taking the appropriate mass limits and substituting the corresponding coupling constants of the GMM or any other  extension model.

\item We corroborated that the $WWV$ amplitude arising from each type of diagrams can be cast in the form of Eq. (\ref{vertex-function}) and  also that all the contributions to the $\Delta\kappa'_V$ and $\Delta Q_V$ form factors are free of ultraviolet divergences.
\end{itemize}

 We now proceed to present the results. Once the amplitude for each  Feynman diagram is written down with the help of the Feynman rules of Fig. \ref{Feynman-Rules},  the Feynman parametrization technique and the Passarino-Veltman method can be applied straightforwardly, followed by some lenghty algebra. Thereafter one can express the  contributions to the $\Delta\kappa'_V$ and $\Delta Q{_V}$ form factors for each type of Feynman diagram of Fig. \ref{Diagrams-abc} as follows

\begin{eqnarray}
\Delta \kappa'^{i}_{V}&=&-\frac{ C^{i}_{V}}{16\pi^2}I^{V-i}_\kappa(x_A,x_B,x_V),\label{Deltakappa-abc}\\
\Delta Q^{i}_{V}&=&-\frac{ C^{i}_{V}}{16\pi^2} I^{V-i}_Q (x_A,x_B,x_V), \label{DeltaQ-abc}
\end{eqnarray}
for $V=Z,\, \gamma$ and $i=a,b,c$. We have introduced the scaled variable $x_I=m_I^2/m_W^2$ ($I=A,\,B$), with $m_A$ and $m_B$ denoting the masses of the  particles circulating into each type of diagram. A word of caution is in order here as $m_A$ and $m_B$, and thereby $x_A$ and $x_B$, are distinct for each type of  contribution. As for the loop functions $I_\kappa^{V-i}$ and $I_Q^{V-i}$, they are presented in Appendix \ref{LoopFunctions}  in terms of parametric integrals and Passarino-Veltman scalar integrals, together with the explicit form of the $C^{i}_{V}$ factors,  which are given in term of the coupling constants of the vertices involved in each Feynman diagram. These coefficients are  presented in Appendix \ref{GMM_Constants} for each possible contribution arising in the GMM.

 As explained above, the  $\Delta \kappa'^{i}_{\gamma}$ and $\Delta Q^{i}_{\gamma}$ form factors can be obtained  from the general expressions  (\ref{Deltakappa-abc})-(\ref{DeltaQ-abc}), and the loop functions presented in Appendix \ref{LoopFunctions},  by taking the $m_V\to 0$ limit. The resulting loop functions $I^{\gamma-i}_{\kappa,\,Q}$ are also  shown in this Appendix. We have verified that these expressions are in agreement with  the results presented in Ref. \citep{Moyotl:2010ss}, where the $WW\gamma$ vertex was studied in the context of little Higgs models.

As far as  the  Feynman diagrams of Fig. \ref{Diagrams-def} are concerned, they only contribute to the $WWZ$ vertex and the respective form factors depend now on three distinct internal masses. They can be written as follows

\begin{eqnarray}
\Delta \kappa'^{i}_{Z}&=&-\frac{ C^{i}_{Z}}{16\pi^2}I^{Z-i}_\kappa(x_A,x_B,x_C,x_Z),\label{Deltakappa-def}\\
\Delta Q^{i}_{Z}&=&-\frac{ C^{i}_{Z}}{16\pi^2} I^{Z-i}_Q (x_A,x_B,x_C,x_Z). \label{DeltaQ-def}
\end{eqnarray}
This time the superscript  $i$ stands for the total contributions of diagrams $i_1$ and $i_2$, with $i=d, e, f$. Expressions for the loop functions in terms of both parametric integrals and Passarino-Veltman scalar integrals can  be found in Appendix \ref{LoopFunctions}.

Once the general expressions for the different kinds of contributions are obtained, we can compute the total contribution of the scalar sector of a given model by simple adding up all the partial contributions. We will present below a numerical analysis of the contributions of the GMM. For the  numerical evaluation we  computed the parametric integrals via the Mathematica  numerical routines. A cross check was done using  the results obtained by evaluating the results given in terms of  Passarino-Veltman scalar functions \cite{Passarino:1978jh} with the help of the LoopTools routines \cite{Hahn:1998yk,vanOldenborgh:1989wn}.

\section{Numerical Discussion}
\label{NumericalDiscussion}
In order to make a numerical evaluation of the contribution of the GMM to the $\Delta\kappa' _V$ and $\Delta Q_V$ form factors,    it is necessary to take into account the current constraints on the parameters space of this model. In particular, our results depend on five free parameters, namely, the singlet mixing angle $\alpha$, the mixing angle between the doublet and the triplet $\theta_H$, and the masses of the new singlet,   $m_H$, the triplet  $m_{H_3}$, and the fiveplet $m_{H_5}$.  A recent study on the indirect constraints on the GMM from $B$ physics and electroweak precision observables can be found in \cite{Hartling:2014aga}, where the   limit on  the triplet VEV $v_\chi\leq 65$ GeV, arising from the measurement of the $b\to s\gamma$ process, was used to impose the strongest bound $\sin \theta_H\leq 0.75$. On the other hand, the current LHC measurements of the couplings and signal strength of the SM-like Higgs boson production \citep{Khachatryan:2014jba, ATLAS:2015bea} constrain in a direct way the  $\theta_H-\alpha$ plane \citep{Chiang:2015amq}.
As for  the masses of the new scalar bosons, experimental constraints on the fiveplet mass have been derived  by the ATLAS collaboration using the like-sign $WWjj$ production  cross-section measurement \citep{Chiang:2014bia}. Furthermore, theoretical constraints from unitarity and vacuum electroweak stability limit the mass of all the scalar bosons of the GMM to be less than 1 TeV  \citep{Chiang:2012cn, Arhrib:2011uy, Aoki:2007ah, Hartling:2014zca}. This constraint was obtained assuming  a $Z_2$ symmetry obeyed by the scalar potential  in order to reduce the number of  free parameters. However, a study  presented in  Ref. \citep{Hartling:2014zca} showed that when the most general potential  (\ref{potencial}) is considered,   there is a decoupling limit in which the masses of the new scalar bosons can be heavy. Therefore, it is interesting  considering the effects when the masses of the new scalar bosons can be heavier than 1 TeV.

\subsection{$\Delta \kappa'_{\gamma}$ and $\Delta Q_{\gamma}$ form factors}

\begin{figure}[htb!]
\begin{center}
\includegraphics[width=17cm]{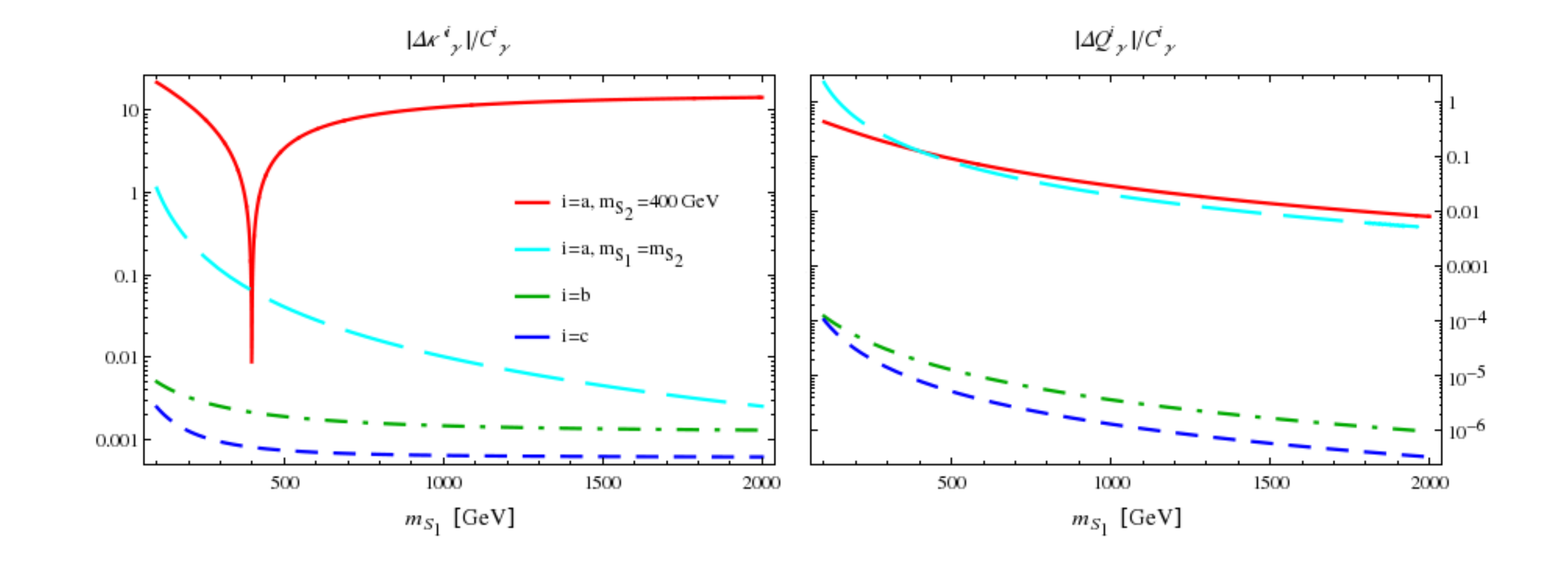}
\caption{Behavior of the contributions of the diagrams of Fig. \ref{Diagrams-abc} to the
$\Delta\kappa'_{\gamma}$ and $\Delta Q_\gamma$ form factors as functions of the masses of the scalar
bosons circulating into the loops of each type of contribution divided by the $C_\gamma^i$
coefficient and in units of $a=g^2/(96\pi^2)$. While type-(a) contribution depends on two scalar
boson masses $m_{S_1}$ and $m_{S_2}$,  type-(b) and type-(c) diagrams depends on  only one scalar
boson mass $m_{S_1}$.
\label{gdeltakappaQ}}
\end{center}
\end{figure}

We list in Tables \ref{typeacont}-\ref{typeccont} of  Appendix \ref{GMM_Constants}  all the  contributions of the GMM to both $\Delta \kappa'_{V}$ and $\Delta Q_{V}$, including the list of  particles circulating into each loop and the explicit form of the corresponding $C_V^i$ coefficient. Excluding the pure SM contributions,  the $\Delta \kappa'_{\gamma}$ and $\Delta Q_{\gamma}$   form factors receive 10 contributions  of  the type-(a) diagrams,  3 of the type-(b) diagrams, and 2 of the type-(c) diagrams. Notice that  all the new scalar bosons participate in the type-(a) diagrams, whereas the type-(b) diagrams only receive contributions from the singlet and the fiveplet scalar bosons, and the type-(c) diagrams from the fiveplet scalar bosons only.  We first examine the general behavior of $\Delta \kappa'_\gamma$ and $\Delta  Q_\gamma$ as functions of the masses of the scalar bosons.   For the type-(b) and type-(c) contributions we show in Fig. \ref{gdeltakappaQ} the form factors  as a function of the mass  of the scalar boson  circulating into the loop, whereas for type-(a) diagram we consider two scenarios: when  both scalar bosons are degenerate and when one scalar boson mass is fixed and the other one is variable.

We first discuss the behavior of $\Delta \kappa'_\gamma$ (left plot of Fig. \ref{gdeltakappaQ}).  As far as type-(a) contribution is concerned, it depends on the masses of two scalar bosons $S_1$ and $S_2$ and is highly dependent on the splitting between their masses $\Delta m_{21}
=m_{S_2}^2-m_{S_1}^2 $.  When such a splitting is  vanishing or very small, $m_{S_2}\simeq m_{S_1}$,
this contribution  decreases quickly as $m_{S_1}$ increases (dashed line), but it   tends to a nonvanishing
constant value when the  splitting becomes large (solid line),
which is in accordance with the decoupling theorem as discussed in Ref.
\cite{TavaresVelasco:2001vb}. It is worth mentioning that the sharp dip observed in the solid line
is due to a change of sign of the form factor, which can become important as there could be large cancellations between contributions due to this change of sign. On the other hand,   the type-(b) and type-(c)
contributions only depend on  one scalar boson mass  and they are larger for a light scalar boson
but decrease quickly when the scalar boson mass increases.  It is important to notice that the
$C_\gamma^{b,c}$ constants are proportional to a factor of  the VEV $v$, thus the size of this type of
 contributions will increase by around two orders of magnitude with respect to the values shown in the
plots.  Even when the scalar boson masses are relatively light, the type-(a) contribution is the
dominant one, except for degenerate masses,  when all the contributions are of similar size.
In summary, the dominant contribution to $\Delta \kappa'_\gamma$ is expected to arise from type-(a)
diagrams, except for a possible suppression due to the $C_\gamma^i$ factor and possible cancellations between distinct contributions. The largest $\Delta \kappa'_\gamma$
value is reached when the scalar boson masses $m_{S_1}$ and $m_{S_2}$ are relatively light or when
there is a large mass splitting $\Delta m_{12}$.

We now turn to analyze the $\Delta Q_\gamma$ form factor, whose dependence on the scalar boson masses is shown in the right plot of Fig. \ref{gdeltakappaQ}. We observe that this form factor exhibits a different behavior to that of $\Delta\kappa'_\gamma$. Although type-(a) contributions are also larger than type-(b) and type-(c) contributions,  in this case there is no dependence on the mass splitting $\Delta m_{21}$ and all the contributions decrease when at least one of the  scalar boson masses becomes large. However, the decrease of $\Delta Q_\gamma$ as $m_{S_1}$ increases is less pronounced that in the case of $\Delta \kappa'_\gamma$. Therefore, barring an extra suppression due to the size of the $C_\gamma^i$ coefficients and possible cancellations, the largest contributions to $\Delta Q_\gamma$ will arise from type-(a) diagrams provided that all the scalar boson masses are lighter. The contribution to this form factor is dominated by the heaviest scalar boson circulating in the type-(a) diagrams and will be very suppressed even  if the other scalar boson is relatively light. In type-(b) and type-(c) diagrams there is also a strong suppression for a heavy scalar boson.

When adding up all the partial  contributions to  $\Delta \kappa'_\gamma$ and $\Delta Q_\gamma$, there could be extra suppression due to the size and sign of the $C_\gamma^i$ coefficients and the loop functions. For instance,  $s_H$ is constrained to be of the order of $10^{-1}$ and thus any  contribution  proportional to this parameter  will have a suppression factor of the order of $10^{-2}$  and  will be  negligible  unless the remaining contributions are also  suppressed. All the contributions of this kind arise from diagrams involving a weak gauge boson and a fiveplet scalar boson. Therefore, all the type-(c) contributions and the type-(b) contributions number 2 and 3  (for the number of each contribution see Table \ref{typeacont} through  Table \ref{typefcont}) will be  two orders of magnitude smaller than the  remaining contributions, although there is a region of the parameter space in which all the contributions are equally suppressed. Even more, the  type-(b) contribution number 1 arises from the loop with the $W$ gauge boson and the $H$ scalar boson, being proportional to the square of the coefficient $f_H=\frac{1}{6}(3c_H s_\alpha-2\sqrt{6} s_H c_\alpha)$, which is very small for small $s_\alpha$ and $s_H$.  Therefore, in most of the allowed region of the parameter space, the largest contributions will arise from the type-(a) diagrams with  two nondegenerate  scalar bosons, though the diagram including the SM Higgs boson and a triplet scalar boson is considerably suppressed as the coefficient $g_h^2$ is very suppressed too. In addition, due to the relative change of sign between distinct  contributions there could be large cancellations once all the type-(a) contributions are added up and so there could be regions of the parameter space where all the three type of contributions are of similar size. However, this region is not the one in which the largest contributions to the form factors can arise.

\begin{figure}[htb!]
\begin{center}
\includegraphics[width=15cm]{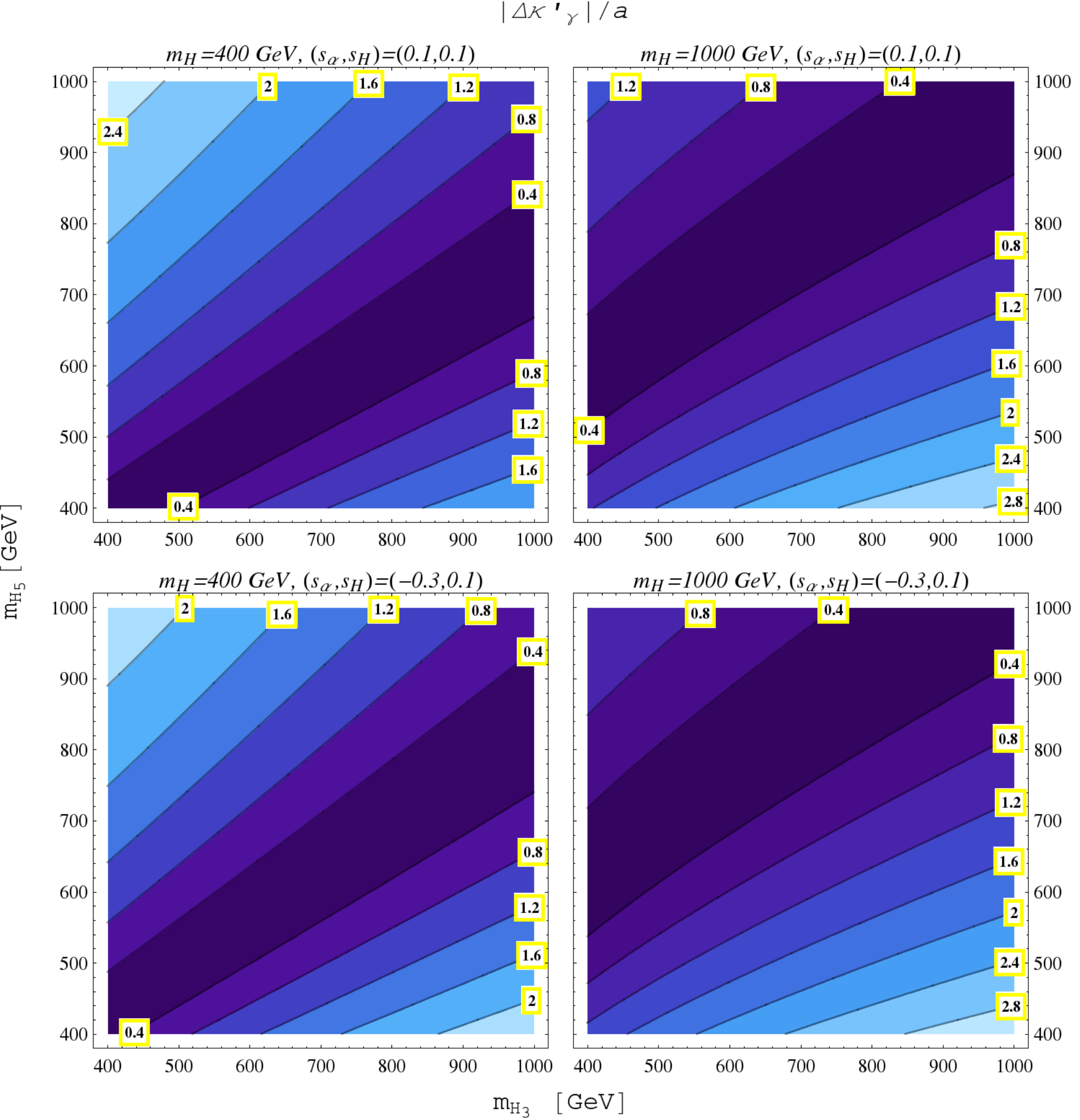}
\caption{Contour plot for the $\Delta\kappa'_{\gamma}$ form factor in the GMM in the $m_{H_3}$ vs $m_{H_5}$ plane for a fixed value of $m_H$ and the indicated values of the mixing angles $s_H$ and $s_\alpha$.
\label{contourkappamH}}
\end{center}
\end{figure}

 All the properties discussed above will reflect on the general behavior of the total contribution from the GMM to the $\Delta \kappa'_\gamma$ and $\Delta Q_\gamma$ form factors, which we have evaluated as  functions of the scalar boson masses. For the mixing angles we used  two combinations of  values lying inside the allowed area of the parameter space   determined  by the authors of Ref. \citep{Degrande:2015xnm} in their  study  of  fiveplet states production at the LHC. We thus considered the sets of values $(s_H,s_\alpha)=(0.1,0.2)$ and $(s_H,s_\alpha)=(0.1,-0.3)$, which  allows us to illustrate the behavior of  $\Delta \kappa'_\gamma$.  As for the masses of the scalar bosons we fix the value of the mass of the singlet  scalar $m_H$ to either 400 and 1000 GeV, and plot in Fig. \ref{contourkappamH} the contour lines of $\Delta \kappa'_\gamma$ in the $m_{H_3}$ vs $m_{H_5}$ plane. In all these plots  the main contributions to $\Delta \kappa'_\gamma$ arise from type-(a) diagrams, though in some regions the type-(b) contributions can be of similar size. We observe that for small $m_H$ (left plots) the largest contributions are reached  for large $m_{H_3}$ and small $m_{H_5}$  and viceversa (lightest area). The region in which $m_{H_3}$ and $m_{H_5}$ are almost degenerate appears in the plots as a dark strip and is the region in which $\Delta\kappa' _\gamma$ reaches its lowest values.     On the other hand, when $m_H$ is large (right plots) we  observe that $\Delta \kappa'_\gamma$ reaches its largest values for large $m_{H_3}$ and light $m_{H_5}$, but in this case there is no such  increase when  $m_{H_5}$ is large and  $m_{H_3}$ remains small, as there are cancellations between the distinct contributions. The dark strip where this form factor reaches its lowest values now has shifted upwards but in general encompasses the area where the three scalar boson masses are large and thereby almost degenerate, namely,  the top right corners of these plots.   We also observe that a change in $s_\alpha$ has a slight impact on the behavior of  $\Delta \kappa'_\gamma$.  However, irrespective of the value of $s_\alpha$, in general the largest values of $\Delta \kappa'_\gamma$ correspond to the scenarios where there is a large splitting between the scalar boson masses and the smallest values correspond to the case when the three masses are large or degenerate.  The largest values of $\Delta \kappa'_\gamma$, in the explored region of the parameter space, are of the order of $a$. In  general the largest contributions arise from type-(a) contributions numbers 2, 4, 5, 7 and 9, but  when all the  masses of the scalar bosons are degenerate these contributions are suppressed and are of similar size than  the type-(b) contribution number 1, which in general is more suppressed than type-(a) contributions.

\begin{figure}[htb!]
\begin{center}
\includegraphics[width=15cm]{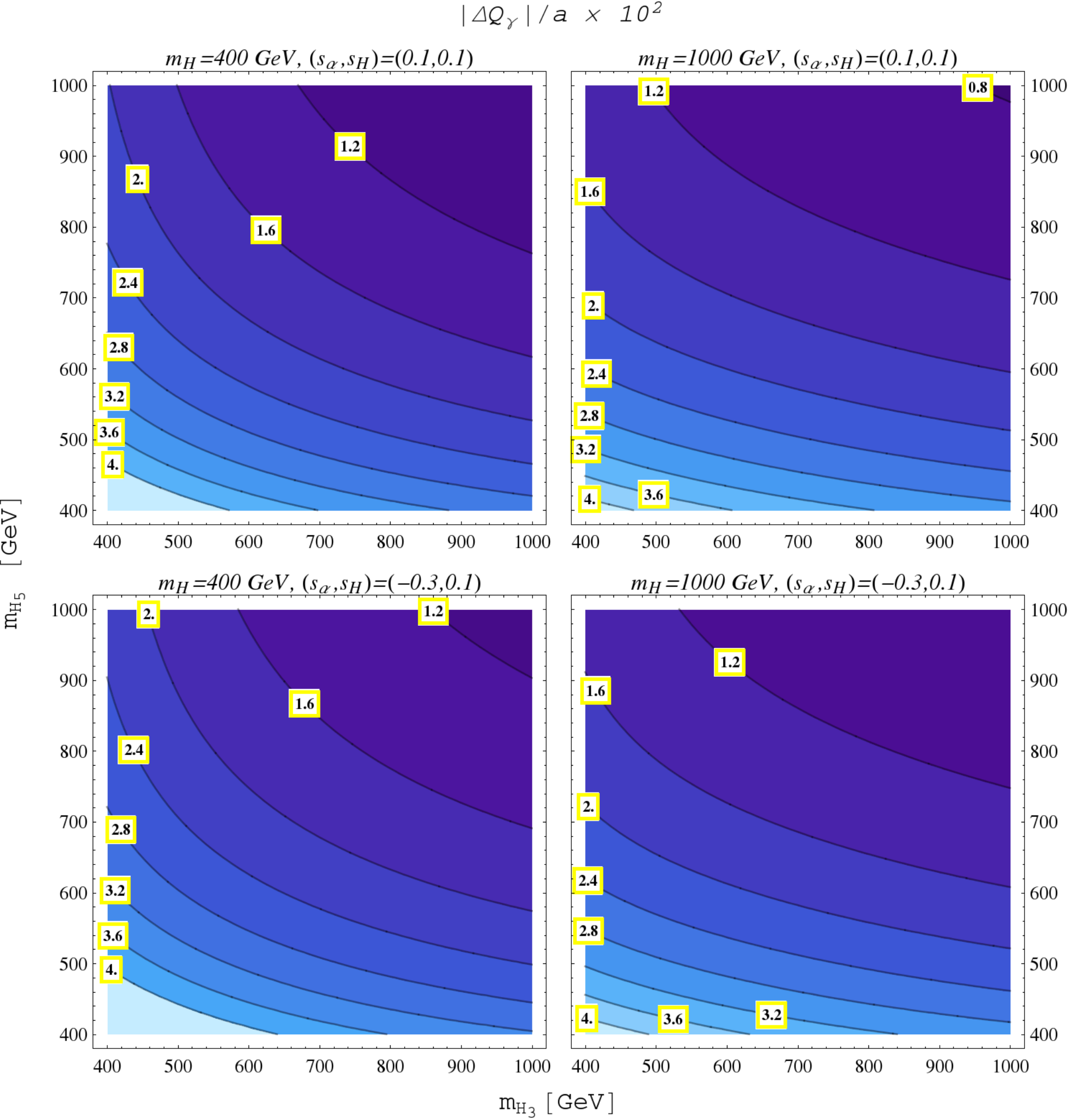}
\caption{The same as in Fig. \ref{contourkappamH}, but for the $\Delta Q_{\gamma}$ form factor.
\label{contourQmH}}
\end{center}
\end{figure}

We now turn to the analysis of the behavior of the $\Delta  Q_\gamma$ form factor. We consider the
same scenarios as in the study of $\Delta\kappa'_\gamma$ and show  in Fig.  \ref{contourQmH} the
contour plot for $\Delta Q_\gamma$ in the $m_{H_5}$ vs $m_{H_3}$ plane.   As discussed above,
contributions of type-(a) have now no dependence on the splitting of the scalar boson masses and they
decrease rapidly as at least one  of the  scalar boson masses becomes large. Therefore,   type-(a)
contributions will reach their largest values in the region (the lightest area)  where  the masses of both
scalars running into the loop  are relatively light.  As for the type-(b) contributions, they have a
similar behavior to type-(a) contributions as they  decrease as the scalar boson mass increases, though in general are smaller
than type-(a) contributions and so are type-(c) contributions. The behavior of the total
contribution  to $\Delta Q_\gamma$ will thus be dominated by the type-(a) contributions and will be
larger  for light degenerate scalar boson masses. This is illustrated in the four plots of Fig.
\ref{contourQmH} in which the largest contributions are reached for small  degenerate masses and they
decrease when either $m_{H_3}$ or $m_{H_5}$ becomes large, though this decrease keeps smooth up to masses of about  800 GeV. In this case the dominant contributions arise from the type-(a) contributions number
6, 8 and 10. When all the masses of the scalar bosons are light, the
type-(a) contribution number 2  is of similar size than contributions 6, 8, and 10, whereas all
other contributions are suppressed due to the small value of the corresponding coefficient
$C_\gamma^a$.  In general, the largest values reached by $\Delta Q_\gamma$  are of the order of one percent of
$a$ and there is a slight dependence  on the value of $s_\alpha$.

It is interesting to note that  the contributions of the GMM to $\Delta\kappa'_\gamma$ are about two orders of magnitude larger than those to $\Delta Q_\gamma$. Such a behavior of the $WW\gamma$ form factors, which  was also  observed for instance  in  the context of a model with   technihadrons \cite{Inami:1995ep} and  the minimal 331 model \cite{TavaresVelasco:2001vb},  can be explained in the light of the decoupling theorem.  It turns out that  $\Delta\kappa'_\gamma$ and $\Delta Q_\gamma$ appear in the $WW\gamma$ vertex function  (\ref{vertex-function}) as coefficients of Lorentz structures of canonical dimension 4 and 6, respectively.  This  means that  $\Delta \kappa'_\gamma$  can be sensitive to nondecoupling effects of  heavy particles, whereas $\Delta Q_\gamma$ is always insensitive to such effects and a natural suppression of this form factor by inverse powers of the mass of the heaviest particle inside the loop is expected. In the present analysis we have considered the contributions of heavy scalar bosons, which explains the observed behavior of the $WW\gamma$ form factors. For a more general discussion of this issue we refer the interesting reader to Refs. \cite{Inami:1995ep, TavaresVelasco:2001vb, Gounaris:1996rz}.   We will see below that, as expected, this feature is also present in the behavior of the $\Delta\kappa'_Z$ and $\Delta Q_Z$ form factors.

\subsection{$\Delta\kappa'_{Z}$ and $\Delta Q_{Z}$ form factors.}
\begin{figure}[htb!]
\begin{center}
\includegraphics[width=17cm]{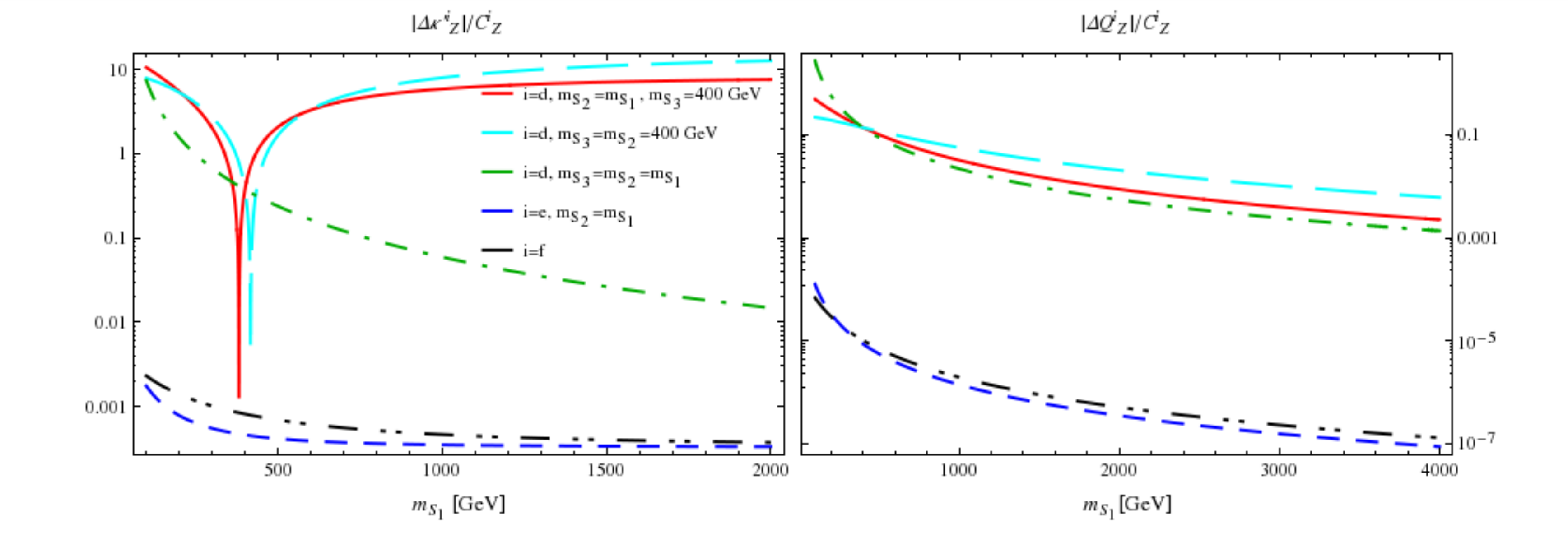}
\caption{Behavior of the contribution of diagrams of Fig. \ref{Diagrams-def} to the
$\Delta\kappa'_{Z}$ and $\Delta Q_Z$ form factors as a function of the masses of the scalar bosons
circulating into the loops of each type of contribution divided by the $C_Z^i$ coefficient.
Type-(d) contribution depends on three scalar boson masses $m_{S_1}$, $m_{S_2}$, and $m_{S_3}$;
type-(e) depends on two scalar masses $m_{S_1}$, and $m_{S_2}$; and type-(f) diagrams depends on
only one scalar boson mass $m_{S_1}$. We only consider the possible scenarios arising in the GMM.
\label{gdeltakappaQZ}}
\end{center}
\end{figure}

We will now  analyze the  $\Delta\kappa'_{Z}$ and $\Delta Q_{Z}$ form factors, for which we  will follow a similar approach to that used above. We thus start by studying the general behavior of the distinct types of contributions. Apart from the diagrams of Fig. \ref{Diagrams-abc}, there is additional contributions due to the diagrams of Fig. \ref{Diagrams-def}. As for the contributions of type (a), (b) and (c), their behavior  is quite similar to that observed in Fig. \ref{gdeltakappaQ}, so  we will  focus on the analysis of the extra contributions, whose behavior will turn out to be  rather similar to that of contributions type (a), (b) and (c). As shown in Appendix \ref{GMM_Constants}, in the GMM there are 7 contributions  of type (d), 4 of type (e), and 3 of type (f). Although our general results allow us to calculate type-(d) contributions with three distinct scalar boson masses $m_{S_1}$, $m_{S_2}$ and $m_{S_3}$, in the GMM all the masses of the same multiplet are degenerate. It means that   type-(d) contributions  arise only from diagrams with at least two degenerate    scalar bosons.  Also, although type-(e) contribution arise from diagrams that can have two distinct scalar bosons,   their masses are degenerate and there is dependence on one mass only, and this is also true for  type-(f) contributions. Therefore, we expect that type-(d) contributions will be the dominant contribution to $\Delta\kappa'_{Z}$ as long as  there is a large mass splitting between the scalar boson masses, whereas type-(e) and type-(f) contributions will only be important for a relatively light scalar boson mass. This is depicted in Fig. \ref{gdeltakappaQZ}, where we show the behavior of the $\Delta\kappa'_{Z}$ and $\Delta Q_{Z}$ form factors for all the scenarios allowed in the GMM. For type-(d) contributions we consider three scenarios:  $m_{S_3}$  fixed and $m_{S_2}=m_{S_1}$  variables, $m_{S_3}=m_{S_2}$  fixed and $m_{S_1}$  variable, and  the three scalar  boson masses  degenerate $m_{S_3}=m_{S_2}=m_{S_1}$.  On the other hand, for type-(e) contributions we only  consider the case when the two scalar bosons are degenerate. In Fig. \ref{gdeltakappaQZ} we observe that $\Delta \kappa'_Z$ and $\Delta Q_Z$ have a similar behavior to that of the $\Delta \kappa'_\gamma$ and $\Delta Q_\gamma$ form factors. In particular, the largest contributions to $\Delta \kappa'_Z$ are reached when there is a large  splitting between the scalar masses or when all the scalar bosons masses circulating into each loop are relatively light. However, the decrease of $\Delta \kappa'_Z$   for large $m_{S_1}$ is now less quick than in the case of $\Delta \kappa'_\gamma$. Again, the $C_Z^i$ factor is proportional to $v$  for type-(e) and type-(f) contributions,  so the values shown in the plots will increase by two orders of magnitude for these contributions. As for $\Delta Q_Z$, it will reach its large value for the smallest allowed scalar boson masses as in the case of $\Delta Q_\gamma$. When the  scalar bosons are very heavy, they will be approximately degenerate, in which case $\Delta Q_Z$ will decrease significantly. Extra suppression for both form factors can arise from the $C_Z^i$ coefficients and from potential cancellations between the distinct contributions as in the case of the electromagnetic form factors.

\begin{figure}[htb!]
\begin{center}
\includegraphics[width=15cm]{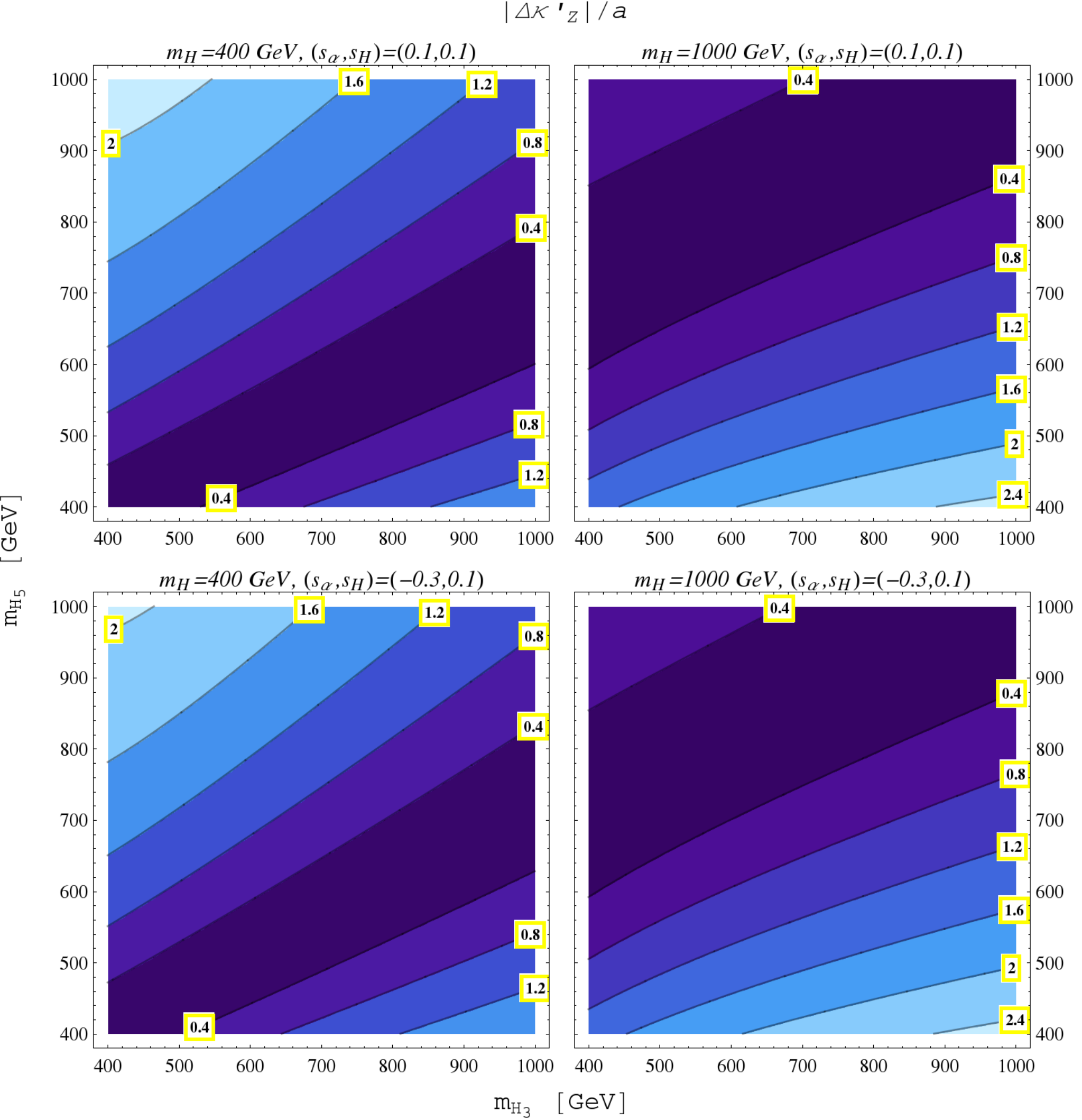}
\caption{Contour plot for the $\Delta\kappa'_Z$ form factor in the GMM in the $m_{H_3}$ vs $m_{H_5}$ plane for a fixed value of $m_H$ and the indicated values of the mixing angles $s_H$ and $s_\alpha$.
\label{contourkappaZmH}}
\end{center}
\end{figure}

In Fig. \ref{contourkappaZmH} we present the contour plots for
 $\Delta \kappa'_Z$ for the same sets of parameter values used above. In spite of the extra
contributions, the behavior of this form factor is rather similar to that of
$\Delta\kappa'_\gamma$. We first note that all the contributions of type (c), (e), and (f) have an
extra suppression due to the $s_H^2$ factor appearing in the respective $C_Z^i$ coefficient and thus
the main contributions will arise from type-(a) and type-(d) contributions, and in lesser extent from
type-(b) contribution number 1. All other contributions  are only important in regions of the
parameter space where the dominant contributions are suppressed by the respective loop
function. As far as the scenario with
$s_\alpha=0.1$ is concerned, we observe in the top left plot, in which we use $m_H=400$ GeV,  that
the largest contributions arise when either $m_{H_3}$ or $m_{H_5}$ are large, whereas in the top right plot we
observe that there is enhancement only when $m_{H_3}$ is large and $m_{H_5}$ remains small, but not in the opposite case. It
means that there are cancellations between contributions when $m_{H_5}$ and $m_H$ are large and thus
the total contribution does not increase in spite of the large splitting between $m_{H_5}$ and
$m_{H_3}$. When the three masses $m_{H}$, $m_{H_3}$, and $m_{H_5}$ are degenerate the total
contribution is suppressed by about one order of magnitude. Even
if all the scalar boson masses are relatively light, $\Delta\kappa'_Z$ is smaller than in the case
where either $m_{H_3}$ or  $m_{H_5}$ are large. In the bottom plots we use $s_\alpha=-0.3$ and observe
that the behavior of $\Delta \kappa'_Z$ has a slight change due to the change in the values of the
$C_Z^i$ coefficients, however its largest values are also of the order of $a$. The darkest
strip where $\Delta \kappa'_Z$ reaches its smallest values, which corresponds to nearly degenerate $m_{H_3}$ and $m_{H_5}$, has now shifted downwards. In summary, the largest values of $\Delta \kappa'_Z$, in this region of the parameter space, are of the order of $a$, and are reached when
there is a large splitting between the masses  scalar bosons. In general the largest contributions to $\Delta \kappa'_Z$ arise from type-(a) and type-(d)
diagrams, with the type (b),(e) and (f) diagrams yielding a subdominant  contribution, which is only relevant when all the masses of the scalar bosons are degenerate.

\begin{figure}[htb!]
\begin{center}
\includegraphics[width=15cm]{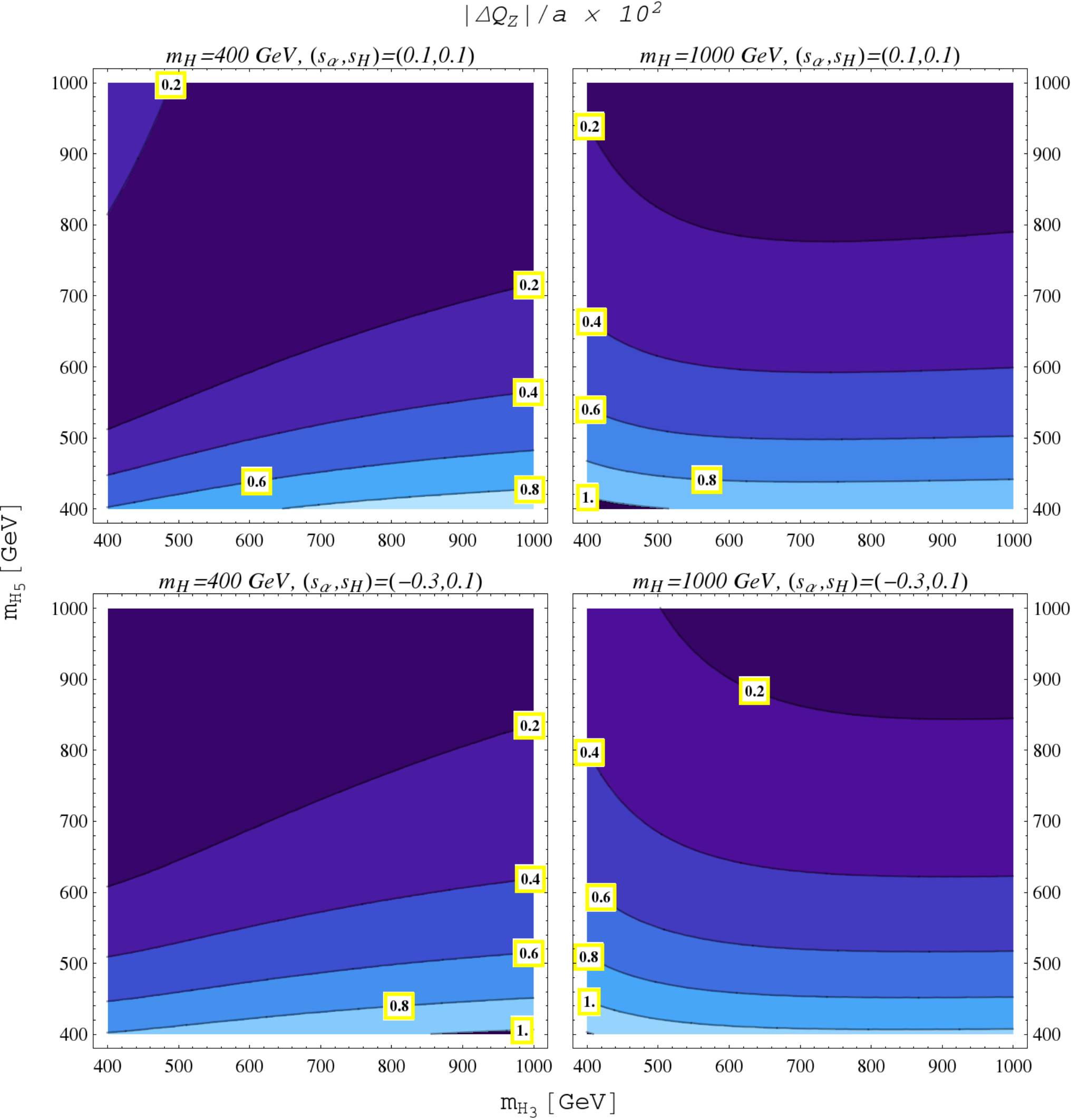}
\caption{The same as in Fig. \ref{contourkappaZmH}, but for the $\Delta Q_Z$ form factor.
\label{contourQZmH}}
\end{center}
\end{figure}

We now turn to the analysis of the behavior of the $\Delta  Q_Z$ form factor, which is shown in Fig.
\ref{contourQZmH}  in the $m_{H_5}$ vs $m_{H_3}$ plane.   As
discussed above, in this case there are no enhancement due to a large splitting of the
scalar boson masses but a decrease when at least one of the masses of the scalar bosons becomes
large. Therefore,  contributions of type-(a) and (d)  reach their largest values provided that all the scalar boson masses are relatively light.  As for the remaining contributions, they have a
similar behavior as they decrease as the scalar boson mass increases, though in general are smaller
than type-(a) and type-(d) contributions. We observe that the largest contributions to $\Delta Q_Z$
arise from diagrams including only fiveplet scalar bosons provided that $m_{H_5}$ is relatively
light irrespective of the value of   $m_H$ and $m_{H_3}$. The behavior of the total
contribution  to $\Delta Q_Z$ is thus  dominated by  type-(a) contributions number 6, 8 and 10, reaching its largest values  for light $m_{H_5}$. Note that type-(a) contributions are the only ones that can involve fiveplet scalar bosons only.
When all the masses of the scalar bosons are light, the type-(a) contributions number 2  and 3 are of
similar size than contributions 6, 8, and 10, whereas all other contributions are suppressed due to
the small value of the corresponding coefficient $C_Z^a$. If $m_{H}$ and $m_{H_3}$ remain small while
$m_{H_5}$ increases there is a cancellation between type-(a) contributions involving singlet and
triplet scalar bosons, such that  the total sum decreases considerably when
$m_{H_5} $ increases.  In general the largest contributions are of
the order of one percent of $a$ in the region of the parameter space considered.

As in the case of the $WW\gamma$ form factors, we also note that the $\Delta \kappa'_Z$ form factor is about two orders of magnitude larger than $\Delta Q_Z$. As it was pointed out above, this behavior can be explained in the context of the decoupling theorem.

\section{Conclusions}
\label{Conclusions}

The presence of new scalars particles is a consequence of  well-motivated extensions of the SM.
Even if such  particles were not  directly produced at particle colliders, their
quantum effects could be at the reach of detection through precision measurement. In this work, we
have obtained the one-loop corrections to the $\Delta\kappa'_{V}$ and $\Delta Q_{V}$ ($V=\gamma,\,Z$) form factors induced by new scalar
particles. A model-independent calculation was done via both the Feynman
parameter technique  and the Passarino-Veltman reduction scheme. Our general results are expressed
in terms of three (six) generic contributions to $\Delta\kappa'_{\gamma}$ and $\Delta Q_{\gamma}$ ($\Delta\kappa'_{Z}$ and $\Delta Q_{Z}$)  that can be used to calculate the corrections arising
from models with an extended scalar sector predicting new neutral, singly, and doubly charged scalar bosons. For the
numerical analysis we have focused on the GMM, which is a Higgs triplet model that has been the source of some
interest recently. This model predicts 9 new scalar bosons accommodated in a singlet, a triplet and
a fiveplet, which yield 15 new contributions to  $\Delta\kappa'_{\gamma}$ and
$\Delta Q_{\gamma}$, whereas $\Delta\kappa'_{Z}$ and $\Delta Q_{Z}$ receive 28  contributions. The general behavior  of the $\Delta\kappa'_{V}$ and $\Delta Q_{V}$  form factors was analyzed for values of the parameters lying inside the region allowed by experimental and theoretical constraints. It was found that  $\Delta\kappa'_{V}$ reaches values  of the order of $a=g^2/(96\pi^2)$, with the largest values arising from the diagrams with two nondegenerate scalar bosons provided that there is a large splitting between their masses. On the other hand $\Delta Q_{V}$ reaches values of the order of one percent of $a$, with the largest contributions arising from diagrams with relatively light degenerate scalar bosons. Both form factors decrease rapidly when all the scalar boson masses are heavy. The values for $\Delta\kappa'_V$ and $\Delta Q_V$ predicted by the GMM are competitive with the ones predicted by other weakly coupled SM extensions, but a very high experimental precision still would be necessary to disentangle such effects.

\subsection{Acknowledgements}
We acknowledge financial support from SNI, CONACYT (M\'exico) and VIEP (BUAP).

\appendix
\section{Feynman rules for the GMM vertices}
\label{Coupling_Constants}
We now present the Feynman rules for the vertices of the type $X_A X_A V$, $\phi_A \phi_B X_C$, and $\phi_A X_B X_C$ arising in the GMM. Here $X$ represents a neutral or charged gauge boson, $V=\gamma, \,Z$, and $\phi$ is a neutral, singly or doubly charged scalar boson. The respective Lorentz structure for each vertex  of this kind was shown in Fig. \ref{Feynman-Rules}, so we only need to present the respective coupling constants. Since in the GMM there is no extra gauge bosons, the only vertices of the type  $X_A X_A V$ are $W^\mp W^\pm \gamma$ and
$W^\mp W^\pm Z$, whose coupling constants are $g_{WW\gamma}=g_\gamma=e$ and $g_{WWZ}=g_Z=gc_W$.
As far as vertices of the class $\phi_A \phi_B X_C$ are concerned, the respective coupling constants are shown in Table \ref{phiAphiBXCconstants}, whereas the coupling constants for vertices of the kind  $\phi_A X_B X_C$ are presented in Table \ref{phiAXBXCconstants}.

\begin{center}
\begin{table}[!htb]
\caption{Coupling constants for vertices of the class $\phi_A \phi_B X_C$ (two scalar bosons and one gauge boson) in the GMM. Here $s_H=\sin\theta_H$ and $c_H=\cos\theta_H$, $g_{h}=\frac{1}{6}\left(2\sqrt{6}c_{H}s_{\alpha}+3s_{H}c_{\alpha}\right)$, and
$g_{H}=\frac{1}{6}\left(2\sqrt{6}c_{H}c_{\alpha}-3s_{H}s_{\alpha}\right)$. For the Lorentz structure see Fig. \ref{Feynman-Rules}.\label{phiAphiBXCconstants}}
\begin{tabular}{ll}
\hline
\hline
Vertex & Coupling constant
\tabularnewline
\hline
\hline
$H_{3}^{\pm}hW^{\mp}$ & $gg_{h}$\tabularnewline
\hline
$H_{3}^{\pm}HW^{\mp}$ & $gg_{H}$\tabularnewline
\hline
$H_{3}^{\pm}H_{5}^{0}W^{\mp}$ & $-\frac{\sqrt{3}}{6}gc_{H}$\tabularnewline
\hline
$H_{5}^{\pm}H_{5}^{0}W^{\mp}$ & $\frac{\sqrt{3}}{2}g$\tabularnewline
\hline
$H_{5}^{\pm}H_{3}^{0}W^{\mp}$ & $\pm\frac{i}{2}gc_{H}$\tabularnewline
\hline
$H_{3}^{\pm}H_{3}^{0}W^{\mp}$ & $\pm\frac{i}{2}g$\tabularnewline
\hline
$H_{5}^{\pm\pm}H_{5}^{\mp}W^{\mp}$ & $-\frac{1}{\sqrt{2}}g$\tabularnewline
\hline
$H_{5}^{\pm\pm}H_{3}^{\mp}W^{\mp}$ & $-\frac{1}{\sqrt{2}}gc_{H}$\tabularnewline
\hline
$H_{3}^{0}hZ$ & $i\frac{g}{c_{W}}g_{h}$\tabularnewline
\hline
$H_{3}^{0}HZ$ & $-i\frac{g}{c_{W}}g_{H}$\tabularnewline
\hline
$H_{5}^{0}H_{3}^{0}Z$ & $-i\frac{g}{\sqrt{3}c_{W}}c_{H}$\tabularnewline
\hline
$H_{5}^{\pm}H_{3}^{\pm}Z$ & $\frac{g}{2c_{W}}c_{H}$\tabularnewline
\hline
$H_{3}^{+}H_{3}^{-}Z$ & $\frac{g}{2c_{W}}(1-2s_{W}^{2})$\tabularnewline
\hline
$H_{5}^{+}H_{5}^{-}Z$ & $\frac{g}{2c_{W}}(1-2s_{W}^{2})$\tabularnewline
\hline
$H_{5}^{++}H_{5}^{--}Z$ & $\frac{g}{c_{W}}(1-2s_{W}^{2})$\tabularnewline
\hline
$H_{3}^{+}H_{3}^{-}\gamma$ & $e$\tabularnewline
\hline
$H_{5}^{+}H_{5}^{-}\gamma$ & $e$\tabularnewline
\hline
$H_{5}^{++}H_{5}^{--}\gamma$ & $2e$\tabularnewline
\hline
\hline
\end{tabular}
\end{table}
\end{center}

\begin{center}
\begin{table}[!htb]
\caption{Coupling constants for vertices of the class $\phi_A X_B X_C$ (one scalar boson and two gauge bosons) in the GMM. Here $f_h=\frac{1}{6}(3c_H c_\alpha+2\sqrt{6} s_H s_\alpha)$ and $f_H=\frac{1}{6}(3c_H s_\alpha-2\sqrt{6} s_H c_\alpha)$. For the Lorentz structure see Fig. \ref{Feynman-Rules}. \label{phiAXBXCconstants}}
\begin{tabular}{ll}
\hline
\hline
Vertex & Coupling constant
\tabularnewline
\hline
\hline
$W^{\pm}W^{\mp}H_{5}^{\pm\pm}$ & $\frac{g^{2}}{\sqrt{2}}vs_{H}$\tabularnewline
\hline
$W^{\pm}ZH_{5}^{\pm}$ & $\mp\frac{g^{2}}{2c_{W}}vs_{H}$\tabularnewline
\hline
$W^{+}W^{-}H_{5}^{0}$ & $\frac{g^{2}}{2\sqrt{3}}vs_{H}$\tabularnewline
\hline
$ZZH_{5}^{0}$ & $-\frac{g^{2}}{\sqrt{3}c_{W}^{2}}vs_{H}$\tabularnewline
\hline
$W^{+}W^{-}h$ & $-g^{2}vf_{h}$\tabularnewline
\hline
$W^{+}W^{-}H$ & $g^{2}vf_{H}$\tabularnewline
\hline
$ZZh$ & $-\frac{g^{2}}{c_{W}^{2}}vf_{h}$\tabularnewline
\hline
$ZZH$ & $\frac{g^{2}}{c_{W}^{2}}vf_{H}$\tabularnewline
\hline
\hline
\end{tabular}
\end{table}
\end{center}

\section{One-loop functions}
\label{LoopFunctions}
In this Appendix we present the results for the loop integrals involved in the $\Delta \kappa'_V$ and $\Delta Q_V$ form factors in terms of parametric integrals and Passarino-Veltman scalar functions.

\subsection{Parametric integrals}

The loop functions arising from the Feynman diagrams of Fig. \ref{Diagrams-abc} can be written in terms of  the following parametric integrals

\begin{equation}
I^{V-i}_{\kappa ,\,Q}=\int_0^1  F^{V-i}_{\kappa ,\,Q} (x)dx,
\label{paramintegral}
\end{equation}
for $V=Z,\gamma$ and $i=a,b,c$. These loop functions depend on $x_A$, $x_B$, and $x_V$, but for the sake of shortness we will drop the explicit dependence from now on. It is worth reminding the reader that subscripts $A$, $B$ correspond to the virtual particles circulating into each Feynman diagram of Fig. \ref{Diagrams-abc}.
 We will first present the  $F^{V-i}_{\kappa,\,Q}(x)$ functions for a massive neutral gauge boson $V$, which can be written as

\begin{eqnarray}
F^{V-i}_\kappa(x)&=& f^{i}_{0}(x)+f^{i}_{1}(x)\tan^{-1}\left[\frac{(x-1)\sqrt{x_V}}{\zeta(x)} \right] +f^{i}_{2}(x)\log[\lambda(x)],
\end{eqnarray}
and
\begin{equation}
F^{V-i}_Q(x)= h^{i}_{0}(x)+h^{i}_{1}(x)\tan^{-1}\left[\frac{(x-1)\sqrt{x_V}}{ \zeta(x)} \right],
\end{equation}
where we introduced the auxiliary function
\begin{eqnarray}
 \zeta(x)&=&\left[4\lambda(x)-(x-1)^2 x_V \right]^{\frac{1}{2}},
\end{eqnarray}
with $ \lambda(x)=x \left(x-\delta-1\right)+x_A$
and $\delta=x_A-x_B$. Also,  $f_j^{i}(x)$ stand for polynomial functions given by
\begin{align}
f_{0}^{a}(x)& =  4\left(x^{2}-1\right),\\
f_{1}^{a}(x)& =  -\frac{4}{\zeta(x)\sqrt{x_{V}}}\left((3x-1)(x-1)^{2}x_{V}+4\lambda(x)(x+1)\right),\\
f_{2}^{a}(x) & =  6x^{2}-8x+2.
\end{align}

\begin{align}
f_{0}^{b}(x) & =  -\frac{1}{2x_{A}^{2}}(x-1)\left(x\left(x_{V}-6x_{A}\right)+x_{V}\right),\\
f_{1}^{b} (x)& =  \frac{1}{2 \zeta(x) x_{A}^{2}\sqrt{x_{V}}}\left(4xx_{V}\left(x\left(x-\delta\right)+\delta_{+}\right)\right.\nonumber\\
 & +  \left.4x_{A}\left(x\left(x\left(7\delta-8x+9\right)-11x_{A}+x_{B}-1\right)+4x_{A}\right)+(x-1)^{2}(3x-1)x_{V}^{2}\right),\\
f_{2}^{b} (x)& =  \frac{1}{4x_{A}^{2}}((4-3x)x-1)x_{V}.
\end{align}

\begin{align}
f_{0}^{c} (x)& =  \frac{1-x^{2}}{x_{B}},\\
f_{1}^{c}(x) & =  \frac{1}{ \zeta(x) x_{B}\sqrt{x_{V}}}\left(4\left(x^{2}-1\right)\left(x-x_{A}\right)+4x(x+3)x_{B}+(3x-1)(x-1)^{2}x_{V}\right),\\
f_{2}^{c} (x)& =  \frac{(4-3x)x-1}{2x_{B}}.
\end{align}
where we have defined $\delta_{\pm }=x_A\pm x_B-1$.

As far as  the polynomial functions  $h_i^j$ are concerned, we only need $h_{i}^{a}$
\begin{align}
h_{0}^{a}(x) & =  -\frac{8(x-1)x}{x_{V}},\\
h_{1}^{a}(x) & =  \frac{32\lambda x}{ \zeta(x) x_{V}^{3/2}},
\end{align}
since the $I_{Q}^{V-b}$ and $I_{Q}^{V-c}$ loop functions obey
 \begin{align}
 \label{I_Q^{V-b}}
I_{Q}^{V-b}&=\frac{2x_A-x_V}{8x_A^2}I_{Q}^{V-a},\\
 \label{I_Q^{V-c}}
I_{Q}^{V-c}&=-\frac{1}{x_B}I_{Q}^{V-a}
.
 \end{align}

As far as the coupling constants  $C^{i}_V$ are concerned, they are as follows
\begin{eqnarray}
\label{CVa}
C^{a}_V&=&\frac{g_{\phi_{A}\phi_{B}W}g_{\phi_{B}\phi_{A}W}g_{\phi_A\phi_A V}}{ g_V}, \\
\label{CVb}
C^{b}_V&=&\frac{g_{X_{A}\phi_{B}W}^2g_{X_AX_A V}}{ m_W^2 g_V},    \\
\label{CVc}
C^{c}_V&=&\frac{g_{\phi_{A}X_{B}W}^2g_{\phi_A\phi_A V}}{ m_W^2 g_V},
\end{eqnarray}
where $g_{ABC}$ stands for the  coupling constants associated with the $ABC$ vertex and presented in Appendix \ref{Coupling_Constants}. Notice that it is necessary to be careful when establishing the  flow of the 4-momenta in the Feynman rule for  each vertex to determine the correct sign of the respective coupling constant.

The  contributions to $\Delta\kappa^{'i} _Z$ and $\Delta Q^i_Z$ from this set of  diagrams   follow easily  after setting  $x_V\to x_Z$ in the above parametric integrals  and inserting the appropriate  coupling constants in the coefficients $C^i_V$ given in Eqs. (\ref{CVa})-(\ref{CVc}). We can also obtain the electromagnetic form factors $ \Delta\kappa'^{i}_{\gamma}$ and $\Delta Q^i_{\gamma}$ straightforwardly by considering the  $x_V\to 0$ limit and the corresponding coupling constants. In this case, the   parametric integrals simplify to

\begin{eqnarray}
I^{\gamma-a}_\kappa&=&2\int_0^1 (x-1)(3x-1)\log \left[ \lambda(x) \right] dx,\label{Fa-gamma}\\
I^{\gamma-b}_\kappa&=&-\int_0^1 \frac{(x-1)^2(x^2+3x_A+\lambda(x))}{2 x_A \lambda(x)}dx ,\label{Fb-gamma}\\
I^{\gamma-c}_\kappa&=&\frac{1}{2} \int_0^1 (x-1) \left[\frac{(1-3 x) \log \left[\lambda (x)
  \right]}{x_B}+\frac{4 x}{\lambda (x) }\right] dx.\label{Fc-gamma}
\end{eqnarray}
and
\begin{equation}
I^{\gamma-a}_Q=\frac{4}{3}\int_0^1  \frac{(x-1)^3 x}{ \lambda(x) }dx ,\label{Ha-gamma}
\end{equation}
with
\begin{eqnarray}
I^{\gamma-b}_Q&=&\frac{1}{4x_A}I^{\gamma-a}_Q,\label{I_Q^{gamma-b}}\\
I^{\gamma-c}_Q&=&-\frac{1}{4 x_B}I^{\gamma-a}_Q.  \label{I_Q^{gamma-c}}
\end{eqnarray}

We now present the parametric integrals for the loop functions of the Feynman diagrams of Fig. \ref{Diagrams-def}, which only contribute to the $\Delta\kappa^{'i} _Z$ and $\Delta Q^i_Z$ form factors. This time the superscript  $i$ stands for whole contribution of diagrams $i_1$ and $i_2$, with $i=d,e,f$.  The parametric integrals $I^{Z-i}_{\kappa,\,Q}$ are given by a similar expression to that of  Eq. (\ref{paramintegral}), but with the $F^{Z-i}_{\kappa,Q}$ functions now depending also on the variable $x_C$. They are  given by

\begin{eqnarray}
F^{Z-i}_\kappa (x)&=& f^{Z-i}_{0}(x)+f^{Z-i}_{1}(x)\eta_{1}(x) +f^{Z-i}_{2}(x)\eta_{2}(x),
\end{eqnarray}
and
\begin{eqnarray}
F^{Z-i}_Q (x)&=& h^{Z-i}_{0}(x)+h^{Z-i}_{1}(x)\eta_{1}(x) +h^{Z-i}_{2}(x)\eta_{2}(x),
\end{eqnarray}
where we introduced the auxiliary functions
\begin{equation}
\eta_{1}(x)= \tan ^{-1}\left[\frac{2(x-1) x_Z}{1+ \delta'^2-(x-1)^2 x^2_Z}\right],
\end{equation}
\begin{equation}
\eta_{2}(x)=\log \left[ \frac{\lambda' (x)}{\lambda(x)}  \right].
\end{equation}
with $\lambda'(x)=x \left(x-\delta'-1\right)+x_C$ and $\delta'=x_C-x_B$.
The $f^{i}_{j}$ and $h^{i}_{j}$ functions are given by

\begin{align}
f_{0}^{d} (x)& =  4(x-1)(3x-1)\log\left[\lambda(x)\right]+8\left(x^{2}-1\right),\\
f_{1}^{d} (x)& =  \frac{4}{ \theta(x) x_{Z}}\left(-2(x+1)x_{Z}\left(-x\left(x_{A}-2x_{B}+x_{C}+2\right)+x_{A}+x_{C}+2x^{2}\right)\right.\nonumber\\
 & +  \left.(5x+1)\delta'^{2}-(x-1)^{2}(3x-1)x_{Z}^{2}\right),\\
f_{2}^{d} (x)& =  \frac{2}{x_{Z}}\left(-(5x+1)x_{A}+5xx_{C}+x_{C}+x(3x-4)x_{Z}+x_{Z}\right).
\end{align}

\begin{align}
f_{0}^{e} (x)& =  -\frac{(x-1)}{x_{C}}\left((3x-1)\log\left[\lambda(x)\right]+2(x+1)\right),\\
f_{1}^{e} (x)& =  \frac{1}{ \theta(x) x_{C}x_{Z}}\left(2x_{Z}\left(x\left(-x\left(x_{A}-2x_{B}+x_{C}\right)+2x_{B}+x_{C}+2x^{2}-2\right)+x_{A}\right)\right.\nonumber\\
 & -  \left.\delta'\left(5xx_{A}+x_{A}-5xx_{C}+x_{C}\right)+(x-1)^{2}(3x-1)x_{Z}^{2}\right)\\
f_{2}^{e} (x)& =  \frac{1}{2x_{C}x_{Z}}\left(5xx_{A}+x_{A}-5xx_{C}+x_{C}+((4-3x)x-1)x_{Z}\right).
\end{align}

\begin{align}
f_{0}^{f}(x)& =  \frac{(x-1)}{2x_{A}x_{B}x_{Z}}\left(2x_{Z}\left(x\left(3x_{A}+9x_{B}-1\right)-3x_{B}-1\right)\right.\nonumber\\
 & +  \left.x_{A}\delta'-(3x-1)\left(3x_{B}+1\right)x_{Z}\log\left[\lambda(x)\right]\right),\\
f_{1}^{f}(x)&=\frac{1}{2 \theta(x) x_{A}x_{B}x_{Z}^{2}}\left(x_{Z}^{2}\left(x^{2}\left(x_{A}\left(8x_{B}+5x_{C}+16\right)+9x_{A}^{2}+22x_{B}\left(x_{C}-2x_{B}\right)+76x_{B}-2x_{C}\right)\right.\right.\nonumber\\
& -4x\left(x_{A}\left(5x_{B}+x_{C}\right)+3x_{A}^{2}+x_{B}\left(-3x_{B}+7x_{C}+2\right)+1\right)\nonumber\\
& \left.-4x^{3}\left(4x_{A}+13x_{B}-1\right)-x_{A}x_{C}+3x_{A}^{2}+6x_{B}x_{C}+2x_{C}\right)\nonumber\\
& +x_{Z}\delta'\left(x_{C}\left(x\left(-4x_{A}-13x_{B}+5\right)-2x_{A}+3x_{B}+1\right)\right.\nonumber\\
& \left.+x_{A}\left(x\left(10x_{A}+7x_{B}-8x+1\right)-4x_{A}+3x_{B}+1\right)\right)\nonumber\\
& \left.+x_{A}\delta'{}^{3}+(x-1)^{2}(3x-1)\left(3x_{B}+1\right)x_{Z}^{3}\right),\\
f_{2}^{f}(x)&=-\frac{1}{4x_{A}x_{B}x_{Z}^{2}}\left(x_{Z}\left(-x_{C}\left(x\left(5x_{A}+13x_{B}-5\right)+x_{A}-3x_{B}\right)\right.\right.\nonumber\\
&+\left.3(3x+1)x_{A}x_{B}+(3x-1)x_{A}\left(3x_{A}-2x-1\right)+x_{C}\right)\nonumber\\
&+\left. x_{A}\delta'^{2}+(x-1)(3x-1)\left(3x_{B}+1\right)x_{Z}^{2}\right),
\end{align}
with
\begin{equation}
\theta(x)=2 x_A \left(x_C-(x-1) x_Z\right)-x_A^2+4 x \left(x_B+x-1\right)
   x_Z-\left(x_C+(x-1) x_Z\right)^2.
\end{equation}

Again we only need the  $h_i^d$ functions
\begin{align}
h_{0}^{d}(x) & =  -\frac{16(x-1)x}{x_{Z}},\\
h_{1}^{d}(x) & =  -\frac{16x\left(\delta'^{2}-x_{Z}\left(x_{A}+x_{C}+2x^{2}-x\left(x_{A}-2x_{B}+x_{C}+2\right)\right)\right)}{ \theta(x) x_{Z}^{2}},\\
h_{2}^{d} (x) & =  \frac{8x\delta'}{x_{Z}^{2}},
\end{align}
whereas the loop functions for the type-(e) and (f) contributions are given by
\begin{align}
 \label{I_Q^{Z-e}}
I_{ Q}^{Z-e}&=-\frac {1}{4x_C} I_{ Q}^{Z-d}\\
 \label{I_Q^{Z-f}}
I_{ Q}^{Z-f}&=\frac {\delta_{+}}{8x_Ax_B} I_{ Q}^{Z-d}
.
 \end{align}

Finally, the $C^i_Z$ coupling constants are

\begin{eqnarray}
\label{CZd}
C^{d}_Z&=&\frac{g_{\phi_{A}\phi_{B}W}g_{\phi_{B}\phi_{C}W}g_{\phi_C\phi_A Z}}{ g_Z}, \\
\label{CZe}
C^{e}_Z&=&\frac{g_{\phi_{A}\phi_{B}W}g_{\phi_{B}X_C W}g_{X_C \phi_{A}Z}}{ m_W^2 g_Z},   \\
\label{CZf}
C^{f}_Z&=&\frac{g_{X_{A}X_{B}W}g_{X_B\phi_C W}g_{\phi_C X_A Z}}{ m_W^2 g_Z}.
\end{eqnarray}

\subsection{Passarino-Veltman scalar integrals}

The  loop functions $I^{V-i}_{\kappa,\,Q}$   were also obtained  via the Passarino-Veltman reduction scheme in terms of two- and three-point scalar functions with the help of the Feyncalc package \cite{Mertig:1990an}.  We first define the following dimensionless ultraviolet finite functions
\begin{align}
\Delta_1&=B_0(0,m_A^2,m_A^2)-B_0(0,m_B^2,m_B^2),\\
\Delta_2&=B_0(m_W^2,m_A^2,m_B^2)-B_0(0,m_B^2,m_B^2),\\
\Delta_3&=B_0(m_V^2,m_A^2,m_A^2)-B_0(m_W^2,m_A^2,m_B^2),\\
\Delta_4&=B_0(0,m_B^2,m_B^2)-B_0(0,m_C^2,m_C^2),\\
\Delta_5&=B_0(m_W^2,m_B^2,m_C^2)-B_0(0,m_C^2,m_C^2),\\
\Delta_6&=B_0(m_V^2,m_A^2,m_C^2)-B_0(m_W^2,m_B^2,m_C^2),\\
\Delta_7&=m_W^2C_0(0,m_W^2,m_W^2,m_A^2,m_A^2,m_B^2),\\
\Delta_8&=m_W^2C_0(m_V^2,m_W^2,m_W^2,m_A^2,m_C^2,m_B^2).
\end{align}
where $B_0(m_i^2,m_j^2,m_k^2)$ and $ C_0(p_1^2,p_2^2,p_{12}^2,m_i^2,m_j^2,m_k^2)$ are two- and three-point scalar functions.

The $I_{\kappa\,Q}^{V-i}$ loop functions can be cast in the following form

\begin{align}
I_{\kappa}^{V-i}&=\frac{1}{D^{V-i}_\kappa}\sum_{j=0}^8 p_{j}^{V-i} \Delta_j+2x_V I_Q^{V-i},\\
I_{Q}^{V-i}&=\frac{1}{D^{V-i}_Q}\sum_{j=1}^8 q_{j}^{V-i} \Delta_j.
\end{align}
with $\Delta_0=1$ and  $i=a,\ldots, f$. For simplicity we have omitted the dependence of the polynomial functions $D_{\kappa',\,Q}^{V-i}$, $p_{j}^{V-i}$, and $q_j^{V-i}$  on $x_A$, $x_B$, and $x_C$.

For the Feynman diagrams of Fig. \ref{Diagrams-abc} we obtain the following polynomial functions for a massive neutral gauge boson $V$

\begin{align}
D_{\kappa'}^{V-a}&=3 y_V^2\\
p_0^{V-a}&=-2 y_V \left(3 \delta^2-x_V+1\right),\\
p_{1}^{V-a}&=-6 x_A y_V \delta_{-}
,\\
p_{2}^{V-a}&=6 \delta y_V \delta_{-}
,\\
p_{3}^{V-a}&=6 \left(6\delta^2-x_A \left(x_V+8\right)+x_B \left(5 x_V-8\right)+x_V+2\right)
,\\
p_{7}^{V-a}&=-12 \left(\rho+x_B x_V\right)\left(3 \delta-x_V+1\right)
,
 \end{align}

 \begin{align}
D_{\kappa'}^{V-b}&=2 x_A^2 y_V^2\\
p_0^{V-b}&=-\frac{1}{6} y_V \left(2 x_A-x_V\right) \left(3 \delta^2-x_V+1\right),\\
p_{1}^{V-b}&=-\frac{1}{2}x_A y_V \delta_{-}\left(2 x_A-x_V\right)
,\\
p_{2}^{V-b}&=\frac{1}{2}\delta y_V \delta_{-} \left(2 x_A-x_V\right)
,\\
p_{3}^{V-b}&=\frac{1}{2}\Big(24 x_A^2 \left(2-x_B-x_V\right)+x_A \left(22 x_B x_V+4 x_B \left(3 x_B-4\right)+5 x_V^2-6 x_V+4\right)\nonumber\\&+12 x_A^3-x_V \left(x_B \left(6 x_B+5 x_V-8\right)+x_V+2\right)\Big)
,\\
p_{7}^{V-b}&=x_A^3 \left(18 x_B+13 x_V-22\right)-3 x_A^2 \left(9 x_B x_V+6 \left(x_B-2\right) x_B+x_V^2+3 x_V-10\right)\nonumber\\&+x_A \left(\left(9 x_B+4\right) x_V^2+\left(x_B \left(17 x_B-16\right)-5\right) x_V+2 \left(3 x_B-1\right) \left(x_B-1\right)^2\right)
\nonumber\\&-6 x_A^4-x_V \left(3 x_B+x_V-1\right) \left(x_B \left(x_B+x_V-2\right)+1\right)
,
 \end{align}

\begin{align}
D_{ \kappa'}^{V-c}&=x_B y_V^2\\
p_0^{V-c}&=\frac{1}{6}y_V \left(3 \delta^2-x_V+1\right),\\
p_{1}^{V-c}&=\frac{1}{2}x_A y_V \delta_{-}
,\\
p_{2}^{V-c}&=-\frac{1}{2}y_V \delta\delta_{-}
,\\
p_{3}^{V-c}&=\frac{1}{2}\Big(x_A \left(12 x_B+x_V+8\right)-6 x_A^2+3 x_B \left(-2 x_B+x_V-8\right)-x_V-2\Big)
,\\
p_{7}^{V-c}&=-x_A^2 \left(9 x_B+x_V+5\right)+x_A \left(x_B \left(9 x_B+x_V+14\right)+2 x_V+1\right)+3 x_A^3\nonumber\\&+x_B \left(-3 x_B \left(x_B+3\right)+\left(x_V-9\right) x_V+11\right)-x_V+1
,
 \end{align}

\begin{align}
D_{Q}^{V-a}&=\frac{3}{4} x_V y_V^3\\
q_0^{V-a}&=y_V \left(12 -2\delta^2 \left(x_V+6\right)+\left(x_V-2\right) x_V\right),\\
q_{1}^{V-a}&=-2 x_A y_V \left(\delta \left(x_V+6\right)-2 \left(x_V+1\right)\right)
,\\
q_{2}^{V-a}&=2 y_V \left(\delta^2\left(x_V+6\right)-2 x_A \left(x_V+1\right)+2x_B \left(2 x_V-3\right)\right)
,\\
q_{3}^{V-a}&=2 \Big(x_A \left(8-x_V \left(3 x_V+20\right)\right)+6 \delta^2 \left(3 x_V-2\right)+3 x_B \left(x_V \left(3 x_V-4\right)+8\right)+2 \left(x_V-1\right) \left(x_V+6\right)\Big)
,\\
q_{7}^{V-a}&=-6 \Big( -2x_Ax_B\delta\left(9 x_V-6\right) -2 x_A^2 \left(x_V \left(x_V+4\right)-2\right)+2 x_A \left(2x_B \left(x_V(2 x_V- 1)+2\right)+x_V \left(2 x_V-1\right)+2\right)\nonumber\\&+x_A^3 \left(6 x_V-4\right)-x_B x_V^3+2 \left(-3 x_B^2+x_B-1\right) x_V^2+2 \left(x_B \left(-3 \left(x_B-2\right) x_B-5\right)+2\right) x_V+4 \left(x_B-1\right)^3\Big)
,
 \end{align}
with $y_V=1-4x_V$, and $\rho=1-2(x_A+x_B)+\delta^2$. Also, the $I_Q^{V-b}$ and  $I_Q^{V-c}$  loop functions obey Eqs. (\ref{I_Q^{V-b}}) and (\ref{I_Q^{V-c}}).

For $V=\gamma$, we need to be careful when taking  the limit $x_V\to 0$ as a result of the form $0/0$ is obtained since the Gram determinant vanishes. Therefore one must recourse to   L'H\^opital rule, as is  described in detail in Ref.  \cite{TavaresVelasco:2001vb}. We obtain  the following results after applying this method

\begin{align}
D_{\kappa'}^{\gamma-a}&=3\\
p_0^{\gamma-a}&=6 \delta^2-3\delta-1,\\
p_{1}^{\gamma-a}&=6 x_A \delta_{-},\\
p_{2}^{\gamma-a}&=6 \left(x_A-\delta^2\right)
,
 \end{align}

\begin{align}
D_{\kappa'}^{\gamma-b}&=2 \rho ^2 x_A\\
p_0^{\gamma-b}&=\frac{1}{6}\Big(\rho  \left(3 \rho  \left(2\delta^2+7x_A +x_B\right)+96 x_A x_B-\rho \right)\Big),\\
p_{1}^{\gamma-b}&=\rho  x_A \left(x_A \left(-4 x_B+\rho -8\right)+4 x_A^2-(\rho +4) x_B-\rho +4\right)
,\\
p_{2}^{\gamma-b}&=-\rho  \left(\rho  \left(\delta^2+3x_A \right)+8 x_A x_B\right)
,
 \end{align}

\begin{align}
D_{\kappa'}^{\gamma-c}&=2 \rho ^2 x_B\\
p_0^{\gamma-c}&=\frac{1}{6}\Big(-\rho  \left(\rho  \left(-3 x_A \left(4 x_B+1\right)+6 x_A^2+3 x_B \left(2 x_B+9\right)-1\right)+48 x_B \delta_{+}\right)\Big),\\
p_{1}^{\gamma-c}&=\rho  x_A \left(1-\delta\right) \left(4 x_B+\rho \right)
,\\
p_{2}^{\gamma-c}&=\rho  \left(\rho  \left(\delta^2-x_A +4x_B \right)+4 x_B \delta_{+}\right)
,
 \end{align}

\begin{align}
D_{Q}^{\gamma-a}&=3 \rho\\
q_0^{\gamma-a}&=-\frac{2}{3}\Big(-3 x_A^3 \left(8 x_B+5\right)+x_A^2 \left(9 x_B \left(4 x_B+3\right)+10\right)-x_A \left(x_B \left(3 x_B \left(8 x_B+3\right)+8\right)-1\right)+6 x_A^4\nonumber\\&+\left(x_B-1\right) \left(6 x_B^3+3 x_B^2+x_B+2\right)\Big),\\
q_1^{\gamma-a}&=-4 x_A\left(\delta-1\right) \left(\left(\delta-1\right)^2-3 x_B\right) ,\\
q_{2}^{\gamma-a}&=4 \left(-\left(4 x_A+1\right) x_B^3+x_A \left(6 x_A-1\right) x_B^2+x_A \left(\left(5-4 x_A\right) x_A-1\right) x_B+\left(x_A-1\right)^3 x_A+x_B^4\right),
 \end{align}
with the $I_{Q}^{\gamma-b}$ and $I_{Q}^{\gamma-c}$ obeying (\ref{I_Q^{gamma-b}}) and (\ref{I_Q^{gamma-c}}).

Finally we present the polynomial functions for the contributions to the  $WWZ$ form factors obtained from the diagrams of Fig. \ref{Diagrams-def}:

\begin{align}
D_{\kappa'}^{Z-d}&=x_Z y_Z^2\\
p_0^{Z-d}&=\frac{1}{3}\Big(-2 y_Z \left(x_Z \left(3 \left(-2 x_A x_B+x_A^2-2 x_B x_C+2 x_B^2+x_C^2\right)+2\right)-6 \left(x_A-x_C\right)^2-2 x_Z^2\right)\Big),\\
p_{1}^{Z-d}&=2 x_A y_Z \left(-x_A \left(x_Z-2\right)+x_B x_Z-2 x_C+x_Z\right)
,\\
p_{2}^{Z-d}&=2 x_Z \left(x_A \left(17 x_B-3 x_C+5\right)-7 x_A^2+x_B \left(-10 x_B+3 x_C+4\right)+7 x_C-2\right)\nonumber\\&+4 \left(x_A-x_C\right) \left(3 x_A-6 x_B+3 x_C+2\right)+2 x_Z^2 \left(\delta^2-4 x_B-1\right)
,\\
p_{4}^{Z-d}&=-2 \Big(-x_A \left(3 x_B \left(3 x_Z-4\right)-5 x_C x_Z+8 x_C+x_Z^2+x_Z+4\right)+x_A^2 \left(x_Z+2\right)\nonumber\\&+x_C \left(x_B \left(-x_Z^2+x_Z-12\right)-7 x_Z+4\right)+x_Z \left(\left(x_B+6\right) x_B x_Z+2 \left(x_B-6\right) x_B+x_Z+2\right)+6 x_C^2\Big)
,\\
p_{5}^{Z-d}&=2 x_Z^2 \left(-x_A-2 x_B \left(x_C-3\right)+x_B^2+\left(x_C-1\right) x_C+1\right)\nonumber\\&+2 x_Z \left(x_A \left(-9 x_B+7 x_C-1\right)+x_A^2+5 x_B x_C+2 \left(x_B-6\right) x_B-6 x_C^2-3 x_C+2\right)\nonumber\\&+4 \left(x_A-x_C\right) \left(x_A+6 x_B-7 x_C-2\right)
,\\
p_{6}^{Z-d}&=2 \Big(x_A \left(2 x_C \left(5 x_Z-8\right)-x_Z \left(12 x_B+x_Z+8\right)\right)+x_A^2 \left(x_Z+8\right)-x_C x_Z \left(12 x_B+x_Z+8\right)\nonumber\\&+2 x_Z \left(x_B \left(6 x_B+5 x_Z-8\right)+x_Z+2\right)+x_C^2 \left(x_Z+8\right)\Big)
,\\
p_{8}^{Z-d}&=-4 \left(3 x_A-6 x_B+3 x_C-2 x_Z+2\right) \Big(-x_Z \left(x_A \left(x_B-x_C+1\right)+x_B \left(-x_B+x_C+2\right)+x_C\right)\nonumber\\&+\left(x_A-x_C\right)^2+x_B x_Z^2+x_Z\Big)
.
 \end{align}

\begin{align}
D_{\kappa'}^{Z-e}&=x_C x_Z^2 y_Z^3\\
p_0^{Z-e}&=\frac{1}{6}\Big(-x_Z y_Z^2 \left(-x_Z \left(3 \left(-2 x_A x_B+x_A^2-2 x_B x_C+2 x_B^2+x_C^2\right)+2\right)+6 \left(x_A-x_C\right)^2+2 x_Z^2\right)\Big),\\
p_{1}^{Z-e}&=\frac{1}{2}\Big(x_A x_Z y_Z^2 \left(x_A \left(x_Z-2\right)-\left(x_B+1\right) x_Z+2 x_C\right)\Big)
,\\
p_{2}^{Z-e}&=-\frac{1}{2}x_Z y_Z\Big(x_A \left(x_B \left(\left(17-2 x_Z\right) x_Z-12\right)-3 x_C x_Z+5 x_Z+4\right)+x_A^2 \left(x_Z-6\right) \left(x_Z-1\right)\nonumber\\&+3 x_C \left(\left(x_B+5\right) x_Z+4 \left(x_B-3\right)\right)+x_Z \left(x_B \left(x_B \left(x_Z-10\right)-4 x_Z+4\right)-x_Z-2\right)-6 x_C^2\Big)
,\\
p_{4}^{Z-e}&=-\frac{1}{2}x_Z y_Z \Big(x_A \left(3 x_B \left(3 x_Z-4\right)-5 x_C x_Z+8 x_C+x_Z^2+x_Z+4\right)+x_A^2 \left(-\left(x_Z+2\right)\right)\nonumber\\&+x_C \left(x_B \left(\left(x_Z-1\right) x_Z+12\right)+15 x_Z-36\right)-x_Z \left(\left(x_B+6\right) x_B x_Z+2 \left(x_B-6\right) x_B+x_Z+2\right)-6 x_C^2\Big)
,\\
p_{5}^{Z-e}&=\frac{1}{2}x_Z y_Z \Big(x_A \left(3 x_B \left(3 x_Z-4\right)-7 x_C x_Z+16 x_C+x_Z^2+x_Z+4\right)+x_A^2 \left(-\left(x_Z+2\right)\right)\nonumber\\&+x_C \left(x_B \left(x_Z \left(2 x_Z-5\right)+12\right)+x_Z \left(x_Z+11\right)-36\right)-x_Z \left(\left(x_B+6\right) x_B x_Z+2 \left(x_B-6\right) x_B+x_Z+2\right)\nonumber\\&-x_C^2 \left(\left(x_Z-6\right) x_Z+14\right)\Big)
,\\
p_{6}^{Z-e}&=\frac{1}{2}x_Z y_Z \Big(x_A \left(x_Z \left(12 x_B+x_Z+8\right)+2 x_C \left(8-5 x_Z\right)\right)+x_A^2 \left(-\left(x_Z+8\right)\right)\nonumber\\&+x_C x_Z \left(12 x_B+5 x_Z-8\right)-2 x_Z \left(x_B \left(6 x_B+5 x_Z-8\right)+x_Z+2\right)-x_C^2 \left(x_Z+8\right)\Big)
,\\
p_{8}^{Z-e}&=-x_Z y_Z \Big(x_A^2 \left(3 x_B \left(x_Z+2\right)-3 x_C x_Z+3 x_C+5 x_Z-2\right)\nonumber\\&+x_A \left(4 x_C \left(3 x_B \left(x_Z-1\right)+\left(x_Z-3\right) x_Z+5\right)-x_Z \left(x_B \left(9 x_B+5 x_Z-2\right)+2 x_Z+1\right)-3 x_C^2 \left(x_Z-1\right)\right)\nonumber\\&-3 x_A^3+x_C x_Z \left(x_B \left(-9 x_B-7 x_Z+10\right)-4 x_Z+7\right)+3 x_C^2 \left(x_B x_Z+2 x_B+3 x_Z-6\right)\nonumber\\&+2 x_Z \left(3 x_B+x_Z-1\right) \left(x_B \left(x_B+x_Z-2\right)+1\right)-3 x_C^3\Big)
.
 \end{align}

\begin{align}
D_{ \kappa'}^{Z-f}&=4 x_A x_B x_Z y_Z^2\\
p_0^{Z-f}&=-\frac{1}{3}y_Z \delta_{+} \Big(x_Z \left(3 \left(-2 x_A x_B+x_A^2-2 x_B x_C+2 x_B^2+x_C^2\right)+2\right)-6 \left(x_A-x_C\right)^2-2 x_Z^2\Big),\\
p_{1}^{Z-f}&=-x_A y_Z \delta_{+}\left(x_A \left(x_Z-2\right)-\left(x_B+1\right) x_Z+2 x_C\right)
,\\
p_{2}^{Z-f}&=x_A^3 \left(x_Z-6\right) \left(x_Z-1\right)-x_A^2 \left(x_B \left(\left(x_Z-10\right) x_Z+6\right)+3 x_Z \left(x_C-x_Z+4\right)-30\right)\nonumber\\&+x_A \left(x_Z \left(x_B \left(7 x_B+64\right)+10 x_C+17\right)-2 \left(x_B \left(40-6 x_C\right)+6 x_B^2+x_C \left(3 x_C+2\right)+18\right)\right.\nonumber\\&-\left.\left(x_B \left(x_B+14\right)+5\right) x_Z^2\right)+\left(x_B-1\right) \left(x_C \left(3 x_B \left(x_Z+4\right)+7 x_Z-4\right)\right.\nonumber\\&-\left.x_Z \left(x_B \left(x_B \left(x_Z-10\right)+4 \left(x_Z-7\right)\right)-x_Z-2\right)-6 x_C^2\right)
,\\
p_{4}^{Z-f}&=x_A^3 \left(-\left(x_Z+2\right)\right)-x_A^2 \left(2 x_B \left(4 x_Z-7\right)+5 x_Z \left(x_Z-x_C\right)+8 x_C-22 x_Z+38\right)\nonumber\\&+x_A \left(x_B \left(x_C \left(\left(x_Z-6\right) x_Z+20\right)+\left(76-17 x_Z\right) x_Z-80\right)+x_B^2 \left(-\left(x_Z-4\right)\right) \left(x_Z-3\right)\right.\nonumber\\&-\left.6 \left(x_C-x_Z\right)^2-12 \left(x_C+3\right)+21 x_Z\right) +\left(x_B-1\right) \left(x_C \left(x_B \left(-x_Z^2+x_Z-12\right)-7 x_Z+4\right)\right.\nonumber\\&+\left.x_Z \left(x_B \left(x_B \left(x_Z+2\right)-2 x_Z+20\right)+x_Z+2\right)+6 x_C^2\right)
,\\
p_{5}^{Z-f}&=x_A^3 \left(x_Z+2\right)+x_A^2 \left(x_B \left(14-8 x_Z\right)+x_C \left(7 x_Z-16\right)+\left(22-5 x_Z\right) x_Z-38\right)\nonumber\\&+x_A \left(x_B \left(-2 x_C \left(\left(x_Z-6\right) x_Z+14\right)+x_Z \left(17 x_Z-76\right)+80\right)+x_B^2 \left(x_Z-4\right) \left(x_Z-3\right)\right.\nonumber\\&+\left.x_C^2 \left(\left(x_Z-6\right) x_Z+14\right)-x_C \left(x_Z \left(x_Z+10\right)-20\right)+3 x_Z \left(2 x_Z-7\right)+36\right)\nonumber\\&+\left(x_B-1\right) \left(-x_C \left(x_B \left(x_Z \left(2 x_Z-5\right)+12\right)+\left(x_Z-1\right) \left(x_Z+4\right)\right)+x_Z \left(x_B \left(x_B \left(x_Z+2\right)-2 x_Z+20\right)+x_Z+2\right)\right.\nonumber\\&+\left.x_C^2 \left(\left(x_Z-6\right) x_Z+14\right)\right)
,\\
p_{6}^{Z-f}&=x_Z^2 \left(x_A \left(21 x_B-x_C+7\right)-5 x_A^2-\left(x_B-1\right) \left(6 x_B+x_C-2\right)\right)\nonumber\\&+x_Z \left(x_C^2 \delta_{+}+2 x_C \left(5 x_A-6 x_B-4\right) \delta_{+}-11 x_A^2 x_B-60 x_A x_B+x_A^3+7 x_A^2\right.\nonumber\\&-\left.4 x_A+12 x_B^3+36 x_B^2-44 x_B\right)+8 \delta_{+}\left(x_A-x_C\right)^2-4 x_Z
,\\
p_{8}^{Z-f}&=-2 \Big(x_A \left(x_C \left(-3 x_B^2 y_Z+x_B \left(x_Z \left(9 x_Z-26\right)+32\right)+x_Z \left(6 x_Z-11\right)+20\right)\right.\nonumber\\&+\left.x_C^2 \left(-9 x_B-8 x_Z+5\right)+x_Z \left(x_B \left(x_B \left(43-13 x_Z\right)+3 x_B^2-x_Z \left(2 x_Z+1\right)+9\right)-6 x_Z+9\right)+3 x_C^3\right)
\nonumber\\&-3 x_A^3 \left(x_Z \left(x_B-x_C+3\right)+x_B+x_C-5\right)+x_A^2 \left(x_B \left(-9 x_C x_Z+9 x_C+7 x_Z^2-40\right)+6 x_B^2 \left(x_Z-1\right)\right.\nonumber\\&-\left.4 \left(x_C-1\right) x_Z^2+3 x_C \left(x_C+3\right) x_Z-x_C \left(3 x_C+17\right)+2 \left(x_Z-9\right)\right)+3 x_A^4\nonumber\\&+\left(x_B-1\right) \left(-x_C^2 \left(3 x_B \left(x_Z+2\right)+5 x_Z-2\right)+x_C x_Z \left(x_B \left(9 x_B+x_Z+14\right)+2 x_Z+1\right)
\right.\nonumber\\&-\left.2 x_Z \left(x_B \left(3 x_B \left(x_B+3\right)-\left(x_Z-9\right) x_Z-11\right)+x_Z-1\right)+3 x_C^3\right)\Big)
.
 \end{align}

\begin{align}
D_{ Q}^{Z-d}&=\frac{3}{8} x_Z^2 y_Z^3\\
q_0^{Z-d}&=y_Z \Big(2 x_A \left(x_B x_Z \left(x_Z+6\right)+x_C \left(16-9 x_Z\right)\right)´+x_A^2 \left(-\left(\left(x_Z-3\right) x_Z+16\right)\right)\nonumber\\&+2 x_B x_C x_Z \left(x_Z+6\right)+x_Z \left(-2 x_B^2 \left(x_Z+6\right)+\left(x_Z-2\right) x_Z+12\right)-x_C^2 \left(\left(x_Z-3\right) x_Z+16\right)\Big),\\
q_{1}^{Z-d}&=-x_A y_Z \left(x_A \left(\left(x_Z-3\right) x_Z+16\right)-x_Z \left(x_B x_Z+6 x_B+2 x_Z+2\right)+x_C \left(9 x_Z-16\right)\right)
,\\
q_{2}^{Z-d}&=-2 x_Z^2 \left(x_A \left(-13 x_B+3 x_C-1\right)+5 x_A^2-3 \left(x_B+4\right) x_C+8 x_B^2+5 x_B+5\right)\nonumber\\&-6 x_Z \left(x_A \left(5 x_B+x_C-9\right)-3 x_A^2-9 x_B x_C+2 x_B^2+x_C \left(4 x_C+9\right)-2\right)\nonumber\\&+36 \left(x_C-x_A\right) \left(x_A-2 x_B+x_C+2\right)+x_Z^3 \left(-2 x_A x_B+x_A^2+x_A+\left(x_B-5\right) x_B-2\right)
,\\
q_{4}^{Z-d}&=x_Z^3 \left(3 x_A+x_B \left(-x_B+x_C-13\right)-2\right)\nonumber\\&+x_Z^2 \left(x_A \left(30 x_B-15 x_C-4\right)-3 x_A^2+2 x_B \left(-10 x_B+4 x_C+17\right)+24 x_C-10\right)\nonumber\\&+2 x_Z \left(x_A \left(23 \left(x_C+1\right)-39 x_B\right)-5 x_A^2+3 \left(x_B \left(5 x_C-8\right)+6 x_B^2-x_C \left(4 x_C+9\right)+2\right)\right)\nonumber\\&+4 \left(x_A-x_C\right) \left(7 x_A+18 x_B-9 x_C-18\right)
,\\
q_{5}^{Z-d}&=x_Z^3 \left(-3 x_A-2 \left(x_B+1\right) x_C+x_B^2+13 x_B+x_C^2+2\right)\nonumber\\&+x_Z^2 \left(x_A \left(-30 x_B+24 x_C+4\right)+3 x_A^2-2 x_B \left(5 x_C+17\right)+20 x_B^2-x_C \left(7 x_C+18\right)+10\right)\nonumber\\&+2 x_Z \left(x_A \left(39 x_B-49 x_C-23\right)+5 x_A^2-3 x_B \left(x_C-8\right)-18 x_B^2+x_C \left(26 x_C+31\right)-6\right)\nonumber\\&-4 \left(x_A-x_C\right) \left(7 x_A+18 x_B-25 x_C-18\right)
,\\
q_{6}^{Z-d}&=x_Z^2 \left(x_A \left(-36 x_B+30 x_C-20\right)+3 x_A^2-12 x_B \left(3 x_C+2\right)+36 x_B^2+x_C \left(3 x_C-20\right)+20\right)\nonumber\\&+2 x_Z \left(2 x_A \left(6 x_B-23 x_C+2\right)+17 x_A^2+4 \left(3 x_B+1\right) x_C-12 \left(x_B-1\right)^2+17 x_C^2\right)\nonumber\\&-64 \left(x_A-x_C\right)^2+x_Z^3 \left(-3 x_A+18 x_B-3 x_C+4\right)
,\\
q_{8}^{Z-d}&=-6 \Big(2 x_Z^3 \left(2 x_A x_B-x_A x_C+x_A+2 x_B x_C-3 x_B^2+x_B+x_C-1\right)\nonumber\\&-x_Z^2 \left(3 x_A^2 \left(x_B-x_C+2\right)+x_A \left(2 x_B \left(6 x_C+1\right)-9 x_B^2-x_C \left(3 x_C+4\right)+1\right)\right.\nonumber\\&+\left.3 \left(x_B+2\right) x_C^2+\left(2-9 x_B\right) x_B x_C+2 x_B \left(3 \left(x_B-2\right) x_B+5\right)+x_C-4\right)\nonumber\\&+2 x_Z \left(-3 x_A^2 \left(x_B+x_C-2\right)+x_A \left(2 x_B \left(6 x_C+1\right)-3 x_B^2-x_C \left(3 x_C+10\right)+1\right)\right.\nonumber\\&+\left.2 x_A^3-3 x_B^2 \left(x_C+2\right)+x_B \left(\left(2-3 x_C\right) x_C+6\right)+2 x_B^3+x_C+2 x_C^2 \left(x_C+3\right)\right)\nonumber\\&-6 \left(x_A-x_C\right)^2 \left(x_A-2 x_B+x_C+2\right)-x_B x_Z^4-4 x_Z\Big)
,
 \end{align}
with the $I_Q^{Z-e}$ and $I_Q^{Z-f}$ functions given by (\ref{I_Q^{Z-e}}) and (\ref{I_Q^{Z-f}}).

\section{$C_V^{i}$ coefficients for all  the new contributions of the GMM to the $\Delta
\kappa'_\gamma$ and $\Delta Q_\gamma$ form factors}
\label{GMM_Constants}

After taking into account all the vertices allowed in the GMM (Appendix \ref{Coupling_Constants}) we can determine the new contributions to the $\Delta \kappa'_V$ and $\Delta Q_V$ form factors arising from the Feynman diagrams of Figs. \ref{Diagrams-abc} and \ref{Diagrams-def}. In   Tables \ref{typeacont} through \ref{typefcont}  we show the  explicit form of the $C_{V}^{i}$ coefficients of Eqs. (\ref{CVa})-(\ref{CVc}) and (\ref{CZd})-(\ref{CZf}) for each such contribution.

\begin{center}
\begin{table}[!htb]
\caption{$C_{V}^{a}$ coefficients for all the type-(a) contributions to the  $\Delta \kappa'_V$ and $\Delta Q_V$ form factors  in the GMM. The second column shows the  particles circulating into the loop and the last two columns show the corresponding $C_{V}^{a}$ factors. \label{typeacont}}
\begin{tabular}{cccc}
\hline
\hline
\#&{\small{}$A\,B$}&$C_{Z}^{a}$&$C_{\gamma}^{a}$\tabularnewline
\hline
\hline
1&{\small{}$H_{3}^{-}h$ } & {\small{}$\frac{g^{2}}{2c_{W}^{2}}g_h^{2}\left(1-2s_{W}^{2}\right)$} & {\small{}$g^{2}g_h^{2}$}\tabularnewline
2&{\small{}$H_{3}^{-}H$} & {\small{}$\frac{g^{2}}{2c_{W}^{2}}g_H^{2}\left(1-2s_{W}^{2}\right)$} & {\small{}$g^{2}g_H^{2}$}\tabularnewline
3&{\small{}$H_{3}^{-}H_{3}^{0}$} & {\small{}$\frac{g^{2}}{8c_{W}^{2}}\left(1-2s_{W}^{2}\right)$} & {\small{}$\frac{g^{2}}{4}$}\tabularnewline
4&{\small{}$H_{3}^{-}H_{5}^{0}$} & {\small{}$\frac{g^{2}c_{H}^{2}}{24c_{W}^{2}}\left(1-2s_{W}^{2}\right)$} & {\small{}$\frac{1}{12}g^{2}c_{H}^{2}$}\tabularnewline
5&{\small{}$H_{5}^{-}H_{3}^{0}$ } & {\small{}$\frac{g^{2}c_{H}^{2}}{8c_{W}^{2}}\left(1-2s_{W}^{2}\right)$} & {\small{}$\frac{1}{4}g^{2}c_{H}^{2}$}\tabularnewline
6&{\small{}$H_{5}^{-}H_{5}^{0}$} & {\small{}$\frac{3g^{2}}{8c_{W}^{2}}\left(1-2s_{W}^{2}\right)$} & {\small{}$\frac{3}{4}g^{2}$}\tabularnewline
7&{\small{}$H_{3}^{+}H_{5}^{++}$} & {\small{}$-\frac{g^{2}c_{H}^{2}}{4c_{W}^{2}}\left(1-2s_{W}^{2}\right)$} & {\small{}$-\frac{1}{2}g^{2}c_{H}^{2}$}\tabularnewline
8&{\small{}$H_{5}^{+}H_{5}^{++}$} & {\small{}$-\frac{g^{2}}{4c_{W}^{2}}\left(1-2s_{W}^{2}\right)$} & {\small{}$-\frac{1}{2}g^{2}$}\tabularnewline
9&{\small{}$H_{5}^{--}H_{3}^{-}$} & {\small{}$\frac{g^{2}c_{H}^{2}}{2c_{W}^{2}}\left(1-2s_{W}^{2}\right)$} & {\small{}$g^{2}c_{H}^{2}$}\tabularnewline
10&{\small{}$H_{5}^{--} H_{5}^{-}$} & {\small{}$\frac{g^{2}}{2c_{W}^{2}}\left(1-2s_{W}^{2}\right)$} & {\small{}$g^{2}$}\tabularnewline
\hline
\hline
\end{tabular}
\end{table}
\end{center}

\begin{center}
\begin{table}[!htb]
\caption{The same as in Table \ref{typeacont}, but for  the type-(b) contributions.\label{typebcont}}
\begin{tabular}{cccc}
\hline
\hline
\#&{\small{}$A B$} & {\small{}$C_{Z}^{b}$} & {\small{}$C_{\gamma}^{b}$}\tabularnewline
\hline
1&{\small{}$W^{-}H$} & {\small{}$-g^{4}\frac{f_H^{2}v^{2}}{m_{W}^2}$} & {\small{}$g^{4}\frac{f_H^{2}v^{2} }{m_{W}^2}$}\tabularnewline
2&{\small{}$W^{-}H_{5}^{0}$} & {\small{}$-\frac{g^{4}s_{H}^{2}v^{2}}{12 m_{W}^2}$} & {\small{}$\frac{g^{4}s_{H}^{2}v^{2}}{12 m_{W}^2}$}\tabularnewline
3&{\small{}$W^{+}H_{5}^{++}$} & {\small{}$\frac{g^{4}s_{H}^{2}v^{2}}{2 m_{W}^2}$} & {\small{}-$\frac{g^{4}s_{H}^{2}v^{2}}{2 m_{W}^2}$}\tabularnewline
\hline
\hline
\end{tabular}
\end{table}
\end{center}

\begin{center}
\begin{table}[!htb]
\caption{The same as in Table \ref{typeacont}, but for  the type-(c) contributions.\label{typeccont}}
\begin{tabular}{cccc}
\hline
\hline
\#&{\small{}${A} {B}$} & {\small{}$C_{Z}^{c}$} & {\small{}$C_{\gamma}^{c}$}\tabularnewline
\hline
\hline
1&{\small{}$H_{5}^{-}Z$} & {\small{}$\frac{g^{4}s_{H}^{2}v^{2}}{8c_{W}^{4}m_{W}^2}\left(1-2s_{W}^{2}\right)$} & {\small{}$\frac{g^{4}s_{H}^{2}v^{2}}{4c_{W}^{2}m_{W}^2}$}\tabularnewline
2&{\small{}$H_{5}^{--}W^{-}$} & {\small{}$-\frac{g^{4}s_{H}^{2}v^{2}}{2c_{W}^2m_{W}^2}\left(1-2s_{W}^{2}\right)$} & {\small{}$-\frac{g^{4}s_{H}^{2}v^{2}}{ m_{W}^2}$}\tabularnewline
\hline
\hline
\end{tabular}
\end{table}
\end{center}

\begin{center}
\begin{table}[!htb]
\caption{$C_{Z}^{d}$ coefficients for  the type-(d) contributions in the GMM. The first column shows the  particles circulating into the loop and the last  column shows the corresponding $c_{Z}^{d}$ factor. \label{typedcont}}
\begin{tabular}{ccc}
\hline
\hline
\#&{\small{}$ABC$} & {\small{}$c_{Z}^{(d)}$}\tabularnewline
\hline
\hline
1&{\small{}$H_{3}^{-} H_{3}^{0} H_{5}^{-}$} & {\small{}$\frac{g^{2}c_{H}^{2}}{8c_{W}^2}$}\tabularnewline
2&{\small{}$H_{3}^{-} H_{5}^{0} H_{5}^{-}$} & {\small{}$\frac{g^{2}c_{H}^{2}}{8c_{W}^2}$}\tabularnewline
3&{\small{}$H_{3}^{+} H_{5}^{++} H_{5}^{+}$} & {\small{}$-\frac{g^{2}c_{H}^{2}}{4c_{W}^2}$}\tabularnewline
4&{\small{}$H_{3}^{0} H_{3}^{+} h$} & {\small{}$\frac{g^{2}}{2c_{W}^2}g_h^{2}$}\tabularnewline
5&{\small{}$H_{3}^{0} H_{3}^{+} H$} & {\small{}$-\frac{g^{2}}{2c_{W}^2}g_H^{2}$}\tabularnewline
6&{\small{}$H_{3}^{0} H_{3}^{+} H_{5}^{0}$} & {\small{}$\frac{g^{2}c_{H}^{2}}{12c_{W}^2}$}\tabularnewline
7&{\small{}$H_{3}^{0} H_{5}^{+} H_{5}^{0}$} & {\small{}$-\frac{g^{2}c_{H}^{2}}{4c_{W}^2}$}\tabularnewline
\hline
\hline
\end{tabular}
\end{table}
\end{center}

\begin{center}
\begin{table}[!htb]
\caption{The same as in Table \ref {typedcont}, but for the type-(e) contributions.\label{typeecont}}
\begin{tabular}{ccc}
\hline
\hline
\#&{\small{}$ABC$} & {\small{}$C_{Z}^{e}$}\tabularnewline
\hline
\hline
1&{\small{}$H_{5}^{-} H_{5}^{0} W^{-}$} & {\small{}$-\frac{g^{4}s_{H}^{2}v^{2}}{8c_{W}^2 m_{W}^2}$}\tabularnewline
2&{\small{}$H_{5}^{+} H_{5}^{++} W^{+}$} & {\small{}$\frac{g^{4}s_{H}^{2}v^{2}}{\sqrt{2}4c_{W}^2 m_{W}^2}$}\tabularnewline
3&{\small{}$H_{5}^{0} H_{5}^{+} Z$} & {\small{}$\frac{g^{4}s_{H}^{2}v^{2}}{4c_{W}^{4} m_{W}^2}$}\tabularnewline
\hline
\hline
\end{tabular}
\end{table}
\end{center}

\begin{center}
\begin{table}[!htb]
\caption{The same as in Table \ref {typedcont}, but for the type-(f) contributions.\label{typefcont}}
\begin{tabular}{ccc}
\hline
\hline
\#&{\small{}$ABC$} & {\small{}$C_{Z}^{f}$}\tabularnewline
\hline
\hline
1&{\small{}$W^{-}ZH_{5}^{-}$} & {\small{}$-\frac{g^{4}s_{H}^{2}v^{2}}{4c_{W}^2 m_{W}^2}$}\tabularnewline
2&{\small{}$ZW^{+}H$} & {\small{}$-\frac{g^{4}v^{2}}{c_{W}^2 m_{W}^2}f_H^{2}$}\tabularnewline
3&{\small{}$Z W^{+}H_{5}^{0}$} & {\small{}$-\frac{g^{4}s_{H}^{2}v^{2}}{6c_{W}^2 m_{W}^2}$}\tabularnewline
\hline
\hline
\end{tabular}
\end{table}
\end{center}



\bibliography{biblio}

\end{document}